\documentclass[preprint]{aastex}
\shorttitle{Revisiting coincidence rate between Gravitational Wave detections and short Gamma-Ray Burst for the Advanced and third generation}
\usepackage{amsmath}

\begin{document}

\title{Revisiting coincidence rate between Gravitational Wave detection and short Gamma-Ray Burst for the Advanced and third generation}
\author{T. Regimbau\thanks{Corresponding author: regimbau@oca.eu}, K. Siellez, D. Meacher, B. Gendre, M. Bo\" er}
\affil{UMR ARTEMIS, CNRS,
University of Nice Sophia-Antipolis, Observatoire de la C\^{o}te d'Azur, CS 34229 F-06304 NICE, France} 

\date{\today}


\begin{abstract}
We use realistic Monte-Carlo simulations including both gravitational-wave and short gamma-ray burst selection effects to revisit the coincident rate of binary systems composed of two neutron stars or a neutron star and a black hole. 
We show that the fraction of GW triggers that can be observed in coincidence with sGRBs is proportional to the beaming factor at $z=0$, but increases with the distance, until it reaches 100 \%  at the GW detector horizon distance. 
When this is taken into account the rate is improved by a factor of $~3$ compared to the simple beaming factor correction. 
We provide an estimate of the performance future GRB detectors should achieve in order to fully exploit the potentiality of the planned third generation  GW antenna Einstein Telescope, and we propose a simple method to constrain the beaming angle of sGRBs.
\end{abstract}


\maketitle


\section{Introduction}\label{Introduction}

The coalescence of compact binary systems, either neutron star - neutron star (BNS) or neutron star - black hole (NS-BH) are among the most promising sources for one of the first direct detections of gravitational waves (GWs) with the new generation of interferometers: the two Advanced-LIGO (aLIGO) \citep{Harry10} in Hanford and Livingston and Advanced-Virgo (AdV) \citep{Acernese06} in Cascina. 
These facilities will be able to detect the late stage of the coalescence, the merger and the ring down of the binary systems located within few hundreds Mpc. The gravitational-wave signal during the adiabatic inspiral phase up to near the last stable orbit and the final damped ring down of the final black hole are accurately described by post-Newtonian expansion and BH perturbation theory, while the progress of numerical relativity over the last decade has provided deep understanding of the merger \citep{Buo09}, giving a good level of confidence for a detection by the network of aLIGO and AdV (in the following we name the combined network consisting of the three interferometers aLIGO and AdV as ALV). 
With the planed third generation GW detector, Einstein Telescope (ET) \citep{Punturo10}, envisioned to consist of three independent V-shaped Michelson interferometers with 60$^\circ$ opening angles, arm lengths of 10 km, arranged in a triangle configuration, and placed underground to reduce the influence of seismic noise, the maximal detection distance is expected to increase significantly over that of ALV, reaching cosmological distances ($z \simeq 4$ for BNS). 

The coalescence of BNS systems is also believed to be at the origin of the short-hard gamma-ray bursts \citep[sGRBs]{Eichler89}. 
In this scenario, the merger of the system produces a transient accretion disk, the gamma-ray emission being produced by the synchrotron and/or inverse Compton scattering from shocks in an ultra relativistic jet \citep[see] [for a review]{Berger13b}. 
Short GRBs might also produce a so-called kilonova through r-processes \citep{Li1998,ros05,Tanvir2005,Berger13a}. 
The standard hypothesis is that the BNS coalescence results in a black hole, though it is possible that a magnetar, or a transient magnetar, is produced as the result of the merger \citep{Usov92,Zhang2006,Corsi2009,Zhang2009}. 

In this context the coincident detection of both GWs and electromagnetic radiation (EM) would be of paramount importance:
\begin{itemize}
\item The coincident detection of a GW and EM event would greatly improve the detection confidence with ALV during the early operations of the facilities.
\item By using an EM detection of a sGRB as a trigger, one can perform a targeted GW search. This would allow for the detection of fainter signals, resulting in a larger horizon distance.
\item On the other hand, GW alerts sent early to GRB satellites and EM telescopes could increase the chance of an EM detection, if the error on sky localization is smaller than the area covered by the satellite \citep{Cannon2012}.
\item Coincident detections could help solve the enigma of GRB progenitors (for instance BNS or NS-BH) but also of the central engine (the GW signature depends on the fate of the system, with the formation of either a black hole or magnetar) and give increased insight in the physics and dynamics of the system
\item GRB with measured redshift, observed also in GW, could be used as standard sirens to constrain the Hubble constant and the dark energy equation of state \citep{Schutz1986,Dalal2006,Nissanke2010,Sathya2010,Zhao2011} or to recover the intrinsic mass distribution by breaking the observed mass-redshift degeneracy.
\end{itemize}

While long GRBs (lGRBs), assumed to originate from massive core collapse supernova have been detected out to a redshift $z= 8.3$ \citep{Zhang2009,Salvaterra09}, the maximal observed distance for sGRBs is considerably closer. 
From a comprehensive search for sGRBs in the rest-frame, \citet{Siellez14} found a maximum distance of $z = 2.74$. 
However, {\it Swift} is not optimal for the detection of sGRBs  whose spectrum is on average harder than that of lGRBs. BATSE and Fermi/GBM may have been able to detect sGRBs out to larger distance, but the larger sky localization errors associated with these experiments prevented any firm association with a given counterpart and thus redshift measurement. 
Hopefully the situation is improving thanks to the intermediate Palomar Transient Factory (iPTF) \citep{Singer13} and other proposed wide-field experiments such as the French ORMES project.

The EM emission of sGRBs (as well as that of lGRBs) is emitted in a narrow beam, though several estimates of the aperture of the jet have been reported \citep{Aasi14}. 
This drastically reduces the chances to have a detection. 

In order to estimate the coincident rate, one has to account for all the selection effects present in both GW and GRB observations. 
This is not trivial as there may be overlaps between them. 
For instance sGRB jets should be directed toward the observer to have a chance to be detected with GRB satellites, but source orientation also affect the strength of the GW signal. 
In this work, we use Monte Carlo simulations that take into account the selection effects of both GW and sGRB observations, to provide realistic estimates on the rate of coincident GW/sGRB detections with both ALV and ET.

The paper is organized as follows: in Section~\ref{sec:MC_simul} we present our simulations, in Section~\ref{sec:efficiency} we derive the coincident efficiency, in Section~\ref{sec:rates} we estimate the coincident rate, in Section~\ref{sec:beaming_angle} we propose a simple method to measure the average value of the sGRB beaming angle, and finally, in Section~\ref{sec:conclusion}  we summarize our main conclusions. 



\section{Monte-Carlo simulations}\label{sec:MC_simul}

\begin{figure}
\includegraphics[trim=0cm 18.5cm 0cm 0cm, width=\columnwidth]{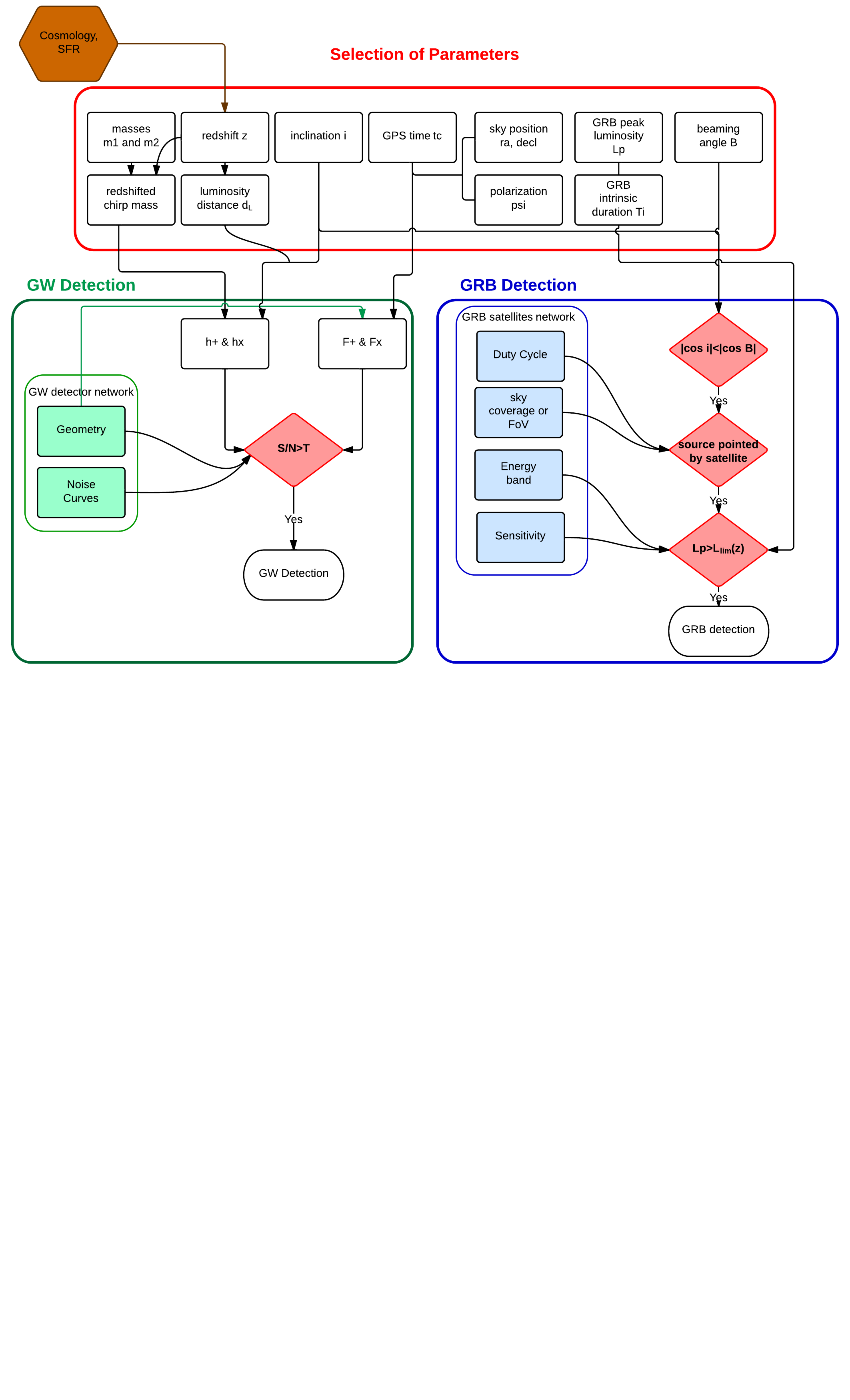}
\caption{A flowchart showing all the different parameters used in these simulations.  
Here the red box indicates the source parameters shown in section~\ref{Simul_BNS}. 
The green box shows the parameters that are used to determine if the event will be observed as a GW detection, see section~\ref{GW_selection}, and the blue box displays the parameters that are used to determine if the event will be observed as a sGRB, see section~\ref{GRB_selection}.
The arrows indicate how the different parameters are related to each other. Colors available on the online version only.
\label{fig:flowchart}}
\end{figure}

In order to investigate the expected rates of coincident detections of both GWs and sGRBs we perform Monte Carlo simulations using distributions in the expected parameter values. Fig.~\ref{fig:flowchart} shows a detailed flow chart of all the source parameters that are used in this work and how they relate to each other. 
We focus here on the population of primordial binaries (pairs of massive stars that survived two core-collapses to form compact systems) and neglect the population of dynamical binaries that may have been formed by captures in dense stellar environment and are not expected to contribute significantly to the total rate \citep{Ivanova08,Sadowski08,Abadie10a}.{\footnote{However, since they have longer evolution time and thus are more numerous at small redshift than primordial binaries, this population may represent the majority of currently observed sGRBs  \citep{Wanderman14}.}} 

\subsection{Simulation of a population of BNS or NS-BH}\label{Simul_BNS}

Short GRBs are thought to be associated in majority with the coalescence of two neutron stars but it has been suggested that the merger of a neutron star and a black hole could also produce a beam of gamma-ray emission. 
Since this scenario cannot be excluded, we consider in this paper separately the two possible sources as the progenitors of sGRBs, binary neutron stars and neutron star-black holes. 
For each population, we first begin by drawing the source parameters, following a procedure similar to that described in \citet{Regimbau12,Regimbau14}.

\begin{figure}
\includegraphics[width=\columnwidth]{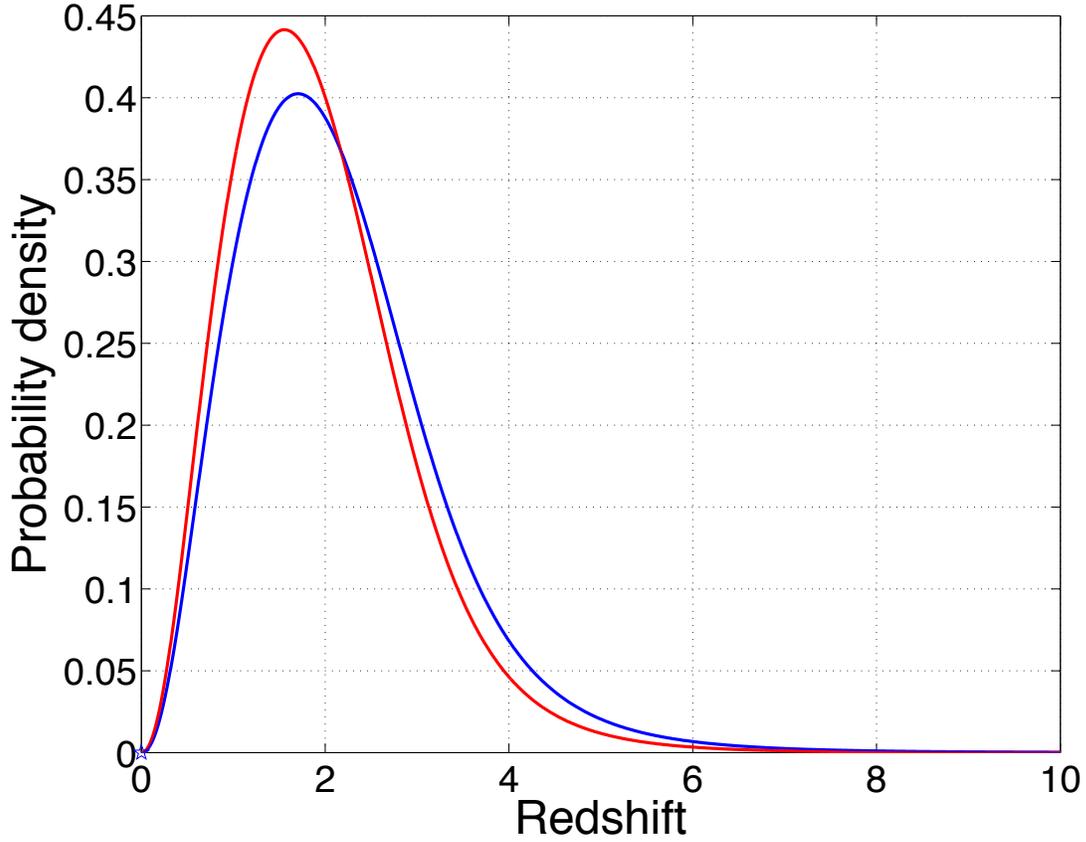}
\caption{Probability distribution of the redshift of BNS (blue, smaller peak) and NS-BH (red, higher peak), assuming the star formation rate of \citet{Hopkins06}, a distribution of the delay of the form $P(t_d) \propto 1/t_d$ with minimal delay of 20 Myr for BNS and 100 Myr for NS-BH. 
\label{fig:Pz}}
\end{figure}

\begin{itemize}
\item The redshift is drawn from a probability distribution $p(z)$ (see in Fig.~\ref{fig:Pz}) constructed by normalizing (in the interval $0-10$) the coalescence rate $\frac{dR}{dz}(z)$, as detailed in Section~\ref{sec:rates}.

\item Each event is given a sky position in equatorial coordinates (declination and right ascension) that is drawn from a isotropic distribution. 
The polarization angle, $\psi$, is selected from a uniform  distribution from [0, $2\pi$]. 
The cosine of the inclination angle, $\cos \iota$, is also drawn from a uniform distribution in the range of [-1, 1].

\item The time interval between two successive events is given by the probability distribution $P(\tau) = exp(- \tau / \lambda)$, assuming coalescences in the observer frame is a Poisson process. 
The average waiting time $\lambda$ is computed from the inverse of the merger rate integrated over all redshifts. 
Equivalently, we can consider that the coalescence time in the observer frame $t_c$ is a uniform distribution in the interval [0, $2\pi$] so as to represent one revolution of the Earth about its axis. 

\item For the initial set of simulations we consider a delta function for the distribution of the masses with $m_{NS} = 1.4 M_\odot$ and $m_{BH} = 10 M_\odot$. 
We will later consider the case of more realistic mass distributions.

\item In order to model the properties of sGRBs, we also set the beaming angle (taken here as the half opening angle of the jet). 
We first investigate fixed angles covering a large range of values in the interval $[5^\circ-30^\circ]$. 
As for the masses, we will consider use of a distribution of angles at a later stage.

\item The intrinsic peak luminosity $L_p$ (in erg s$^{-1}$) is drawn from the standard broken power law distribution first proposed by \citet{Guetta05}:
\begin{equation}
\Phi(L_p) \propto
\left\{
\begin{aligned}
(L_p / L_*)^{\alpha} &\,\ & \mathrm{if} & \,\   L_* / \Delta_1< L_p < L_* , \\  
(L_p / L_*)^{\beta}  &\,\ &\mathrm{if} & \,\  L_*< L_p < \Delta_2 L_* ,
\end{aligned}
\right.
\end{equation}
with $\alpha=-0.6$, $\beta=-2$. We adopt the value of $L_*= 10^{51}$ erg s$^{-1}$, corresponding to model ii of \citet{Guetta05b,hopman05,Guetta09} for primordial binaries, to which we have applied a factor of $\sim 1/2$ correction to convert $L_*$ in the band $50-300$ keV to the band $15-150$ keV used in this paper (see Section~\ref{GRB_selection}). 
This is in agreement with the recent work of \citet{Wanderman14}{\footnote{actually there is a factor $\sim 20$ difference  in the results of \citet{Wanderman14} since they used a larger energy band ($1-10000$ keV)}}. 
We also consider a conservative value of  $L_*=  5 \times 10^{50}$ erg s$^{-1}$ (the lower bound of model ii of \citet{Guetta05b}), which accounts for a possible extra bias arising if, among the observed sample of sGRBs, those with redshift measurement are the most luminous. 
We choose $\Delta_1=100$ and $\Delta_2=10$, in order to cover more than $99\%$ of the luminosities. 
Notice that taking $\Delta_1=30$ as suggested by \citet{Guetta05b} has a very small effect and does not affect the final results.
We neglect in this work any possible evolution of the luminosity with redshift \citep{Butler10,Howell14}.

\item The log of the intrinsic duration of the burst, $\log T_i$, is drawn from a Gaussian distribution of mean $\mu_{\log T_i} =-0.458$ and standard deviation $\sigma_{\log T_i} = 0.502$, derived by fitting the sample of \citet{Zhang12}. 
We neglect here any possible correlation between the peak luminosity and the duration.

\end{itemize}


\subsection{GW selection effects}\label{GW_selection}

For each coalescence we must first determine if its resultant GW emission is detectable. 
For this purpose we calculate the event's coherent signal-to-noise ratio (SNR), for the detector network, in the ideal case of Gaussian noise (see \citet{Ghosh13} for a more sophisticated scenario including the possibility of false alarms).

\begin{figure}
\includegraphics[width=\columnwidth]{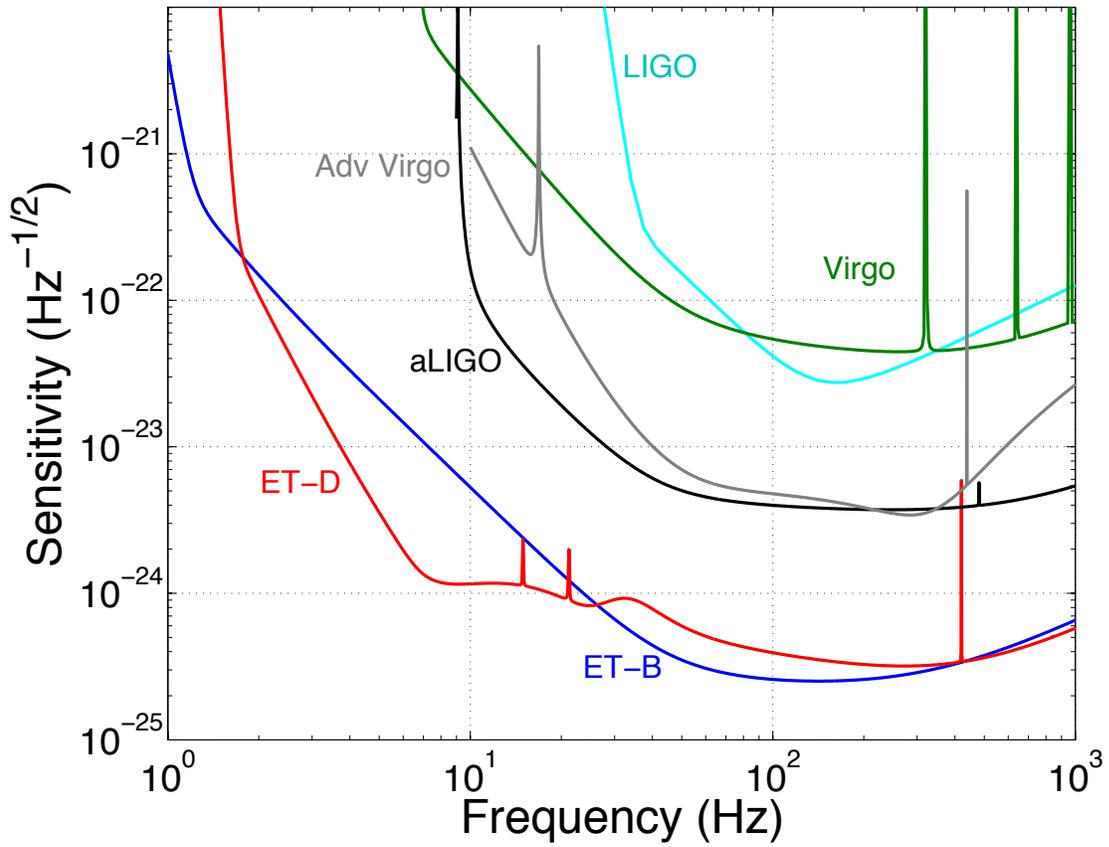}
\caption{Projected sensitivity for second generation (advanced) detectors (here the aLIGO high-power zero detuning sensitivity \citep{Harry10} and Adv Virgo BNS optimized \citep{Acernese06}) and for the initial configuration of ET, ET-B, considered in the Design Study, and the most evolved configuration ET-D \citep{Punturo10}. The sensitivity of first generation detectors LIGO and Virgo is also shown for comparison.} 
\label{fig:sensitivity}
\end{figure}

The SNR detected by matched filtering with an 
optimum filter, in a detector labelled $A$, is: 
\begin{equation}
 \rho_A^2 = 4 \int_0^\infty \frac{|\tilde{h}_{+}F_{+,A} +\tilde{h}_{\times} F_{\times,A}|^2}{S_{n,A}(f)}\, {\rm d}f,
\end{equation}
where $f$ is the frequency of the gravitational wave in the observer frame, $\tilde{h}_{+}$ and $\tilde{h}_{\times}$ the Fourier transforms
of the GW strain amplitudes of polarisations + and $\times$, $F_{+,A}$ and $F_{\times,A}$ the antenna response functions to the GW polarisations, and $S_{n,A}(f)$ the one-sided noise power 
spectral density (PSD) of detector A (see Fig.~\ref{fig:sensitivity}).

For low mass systems such as BNS or NS-BH, the SNR is dominated by the inspiral part of the signal and can reduce to:
\begin{equation}
\rho_A^2 = \frac{5}{6} \frac{(G\mathcal{M}(1+z))^{5/3} {\cal F}_A^2}{c^3 \pi^{4/3} d_L^2(z)} \int_{f_{\mathrm{min}}}^{f_{LSO}(z)} df \,  \frac{f^{-7/3}}{S_{n,A}(f)} .
\label{eq:SNR}
\end{equation}
\noindent Here $\mathcal{M}$ is the intrinsic chirpmass, a combination of the two component masses, $d_L(z)$ is the luminosity distance, $G$ is the gravitational constant, $c$ is the speed of light, $f_{\min}$ is the low frequency limit of the detector and $f_\mathrm{LSO}(z)=f_\mathrm{LSO}/(1+z)$ is the observed (redshifted) gravitational-wave frequency of the last stable orbit.
The factor:
\begin{equation}
{\cal F}_A^2 = \frac{(1+\cos^2 \iota)^2}{4}  F_{+,A}^2 + \cos^2 \iota \, F_{\times,A}^2 ,
\end{equation}
\noindent characterises the detector response.
Assuming uncorrelated noise, the combined SNR for the network of detector is simply the quadrature sum $\rho^2=\sum \rho^2_A$ of individual SNRs. 
If $\rho$ is larger than a set SNR  threshold level ($\rho \ge \rho_\mathrm{T}$) then we say that the event is detectable. 


\subsection{EM selection effects}\label{GRB_selection}

\subsubsection{The beaming angle}

The first selection effect affecting the detection of sGRBs in EM is the strong focussing of the ultra-relativistic jetted emission, i.e.  the beaming angle $\theta_B$. 
Only the fraction $\Theta_B = (1-\cos \theta_B)$ of sources with inclination angle:
\begin{equation}
| \cos \iota | \leq \cos \theta_B \,\ (0 \leq \iota \leq \theta_B \,\ \mathrm{or} \,\ \pi \leq \iota \leq\pi+\theta_B) ,
\label{eq-SE_angle}
\end{equation} 
can be observed on Earth.

Because of the small number of detections of afterglow associated with sGRBs, this angle is constrained to $5^{\circ} - 10^{\circ}$ for only a handful of cases \citep[see eg][]{Soderberg06, Fong12, Nicuesa11}. 
The non detection of a jet break also provides lower limits on the jet opening angle between 5$^\circ$ and 25$^\circ$ \citep{Coward12, Berger13b, Fox06, Grupe06}. 
Based on a sample of 79 lGRBs and 13 sGRBs \citep{Fong12} propose a median of about 10$^\circ$. 
Because of this uncertainty we use a wide range of values for the jet opening angle between 5$^\circ$ - 30$^\circ$.

\subsubsection{Instrumental effects}

The other obvious selection effects are related to the instrument: the burst can be undetected due to its faintness, to the fact that it is not located in the Field Of View (FOV), or because it occurs while the instrument cannot record it. This last effect is dominated by passes through the Southern Atlantic Anomaly (SAA), which reduce the Duty Cycle (DC) to $\sim 80 \%$ of the total time. In the following, we take the effects of the FOV and the DC as independent factors reducing the probability of detection, independently of the source distance and spectrum.
Table \ref{tab:detect} gives a summary of the characteristics of the detectors we have considered in this work.

The trigger conditions can be complex, either for Fermi or \textit{Swift}. For Fermi, GBM triggers when two or more detectors exceed background by n sigma over t timescale in e energy band. To do so, 62 algorithms operate simultaneously with different value for the threshold above background, various timescales and energy bands. Moreover, the energy and sensitivity threshold have been modified 8 times over 4 years as reported in the $2^{nd}$ Fermi-GBM GRB catalog~\citep{Kien14}. This resulted in a 20\% difference in sGRB triggers over the period. For the \textit{Swift} satellite, based on the photon count rates, the complex trigger algorithm adopted uses hundreds of criteria to maximize the GRB observations and cover the largest range of possible burst duration. Each criterion uses different time ranges for the foreground periods and backgrounds, to find the accurate SNR. If this SNR passes the threshold, then the event will have to pass a slew of procedures to confirm its nature. Because of the relative stability of the \textit{Swift} trigger conditions, we decided to use it as a reference for the sensitivity. A simple scaling relation can be use to relate Swift and Fermi rates.

In this work, we restrict the trigger selections to the condition:

\begin{equation}
L_p \geq  L_{\mathrm{lim}}(z) \,\ \mathrm{with} \,\ L_{\mathrm{lim}}(z)= F_{\mathrm{lim}} 4 \pi d_L(z)^2 k(z),
\label{eq-SE_flux}
\end{equation}
\noindent where $L_p$ is the peak luminosity of a source at redshift $z$, $F_{\mathrm{lim}}$ is the limiting flux for a sGRB detection, and $k(z)$ is the k-correction due to the finite observation band at a given redshift.
The limiting flux depends on the spectral properties of each trigger and also on whether it is seen on-axis or off-axis. For a generic sGRB defined by its peak energy $E_p = 440$ keV and low and high energy spectrum power indices $\alpha = -0.5$, and $\beta = -3.2$, the Band law \citep{Band93} gives $F_{\mathrm{lim}} \sim 0.4$ ph s$^{-1}$ cm$^{-2}$ for on-axis triggers in the $1-1000$ keV observation band. A moderate variation of $E_p$ (up to a factor $2-3$) does not change significantly this result. For the {\em Swift} detection band $15-150$ keV, it gives $F_{\mathrm{lim}} \sim 0.56$ ph s$^{-1}$ cm$^{-2}$, which translates to $F_{\mathrm{lim}} \sim 1.5$ ph s$^{-1}$ cm$^{-2}$ for off-axis sources. In addition to these optimal and averaged values, we also consider a pessimistic value of 2.5 ph s$^{-1}$ cm$^{-2}$, corresponding to sGRBs with redshift measurement.

\begin{figure}
\includegraphics[width=\columnwidth]{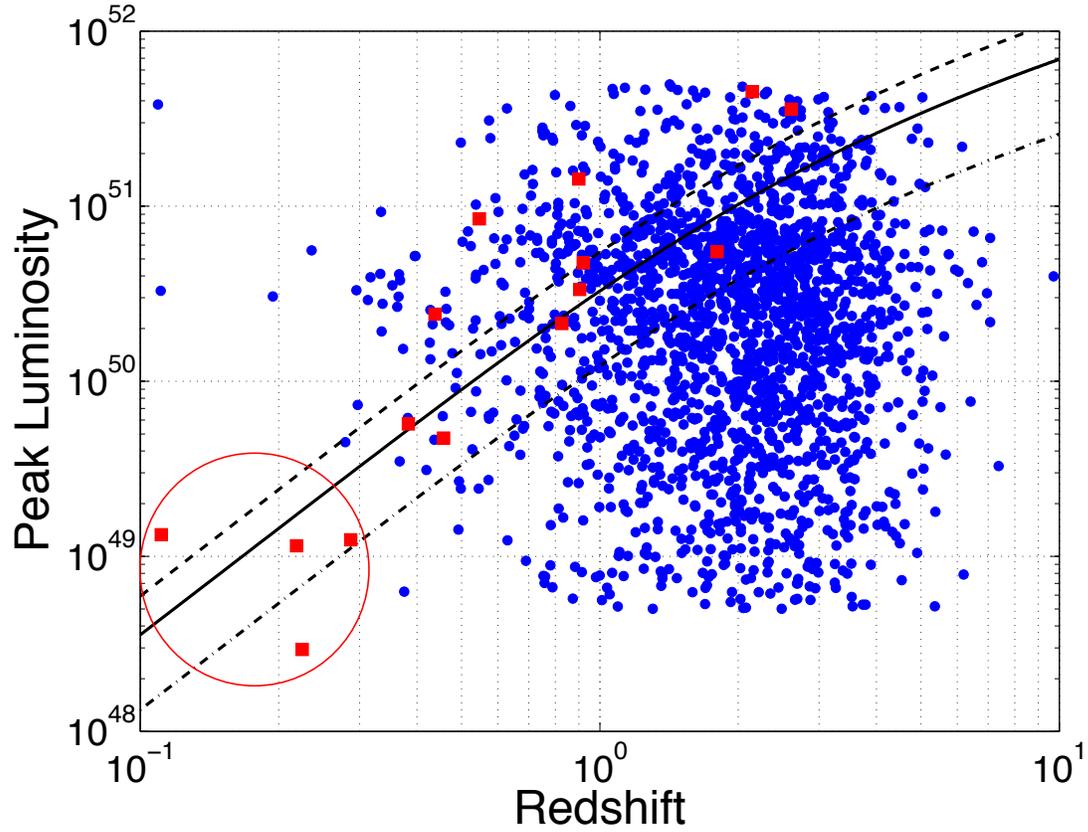}
\caption{Sensitivity of the \textit{Swift} satellite for a flux limit of 1.5 ph s$^{-1}$cm$^{-2}$ (continuous black line), 2.5 ph s$^{-1}$cm$^{-2}$ (dashed black line), corresponding to sGRBs with redshift measurement, and an optimal value of 0.56 ph s$^{-1}$cm$^{-2}$ (dash-dotted black line), corresponding to on-axis sources. The blue circles correspond to a sample of 10000 sources simulated from the Monte Carlo procedure described in \ref{sec:MC_simul}. The red squares show a sample of 17 observed  sGRBs \citep{Zhang12}. The low redshift/low luminosity population, indicated by a red circle in the bottom left of the plot, is difficult to reproduce with the simulations and could be either a population of magnetars or dynamical BNS or NS-BH whose long evolution times, of the order of 3 Gyr \citep{Wanderman14}, may explain why they are more numerous at low redshift. It would not explain why they would be sub-luminous though.}
\label{fig:sens_Swift}
\end{figure}

Fig.~\ref{fig:sens_Swift} shows $L_{\mathrm{lim}}(z)$ for these three limits, along with a sample of 17 observed sGRBs with redshift measurement \citep{Zhang12}, and 1000 simulated data with the conservative peak luminosity distribution ($L_*=5 \times 10^{50}$), which best fit the observed sample (using $L_*=10^{51}$ produces too many detected sGRBs at large redshift).
\section{Detection efficiency}\label{sec:efficiency}

In the following sections, we present the various selection effects described above in terms of efficiency, i.e. the fraction of sources, relative to the total, which can be detected at a given redshift. 
\subsection{GW efficiency}

\begin{figure}
\includegraphics[width=0.49\columnwidth]{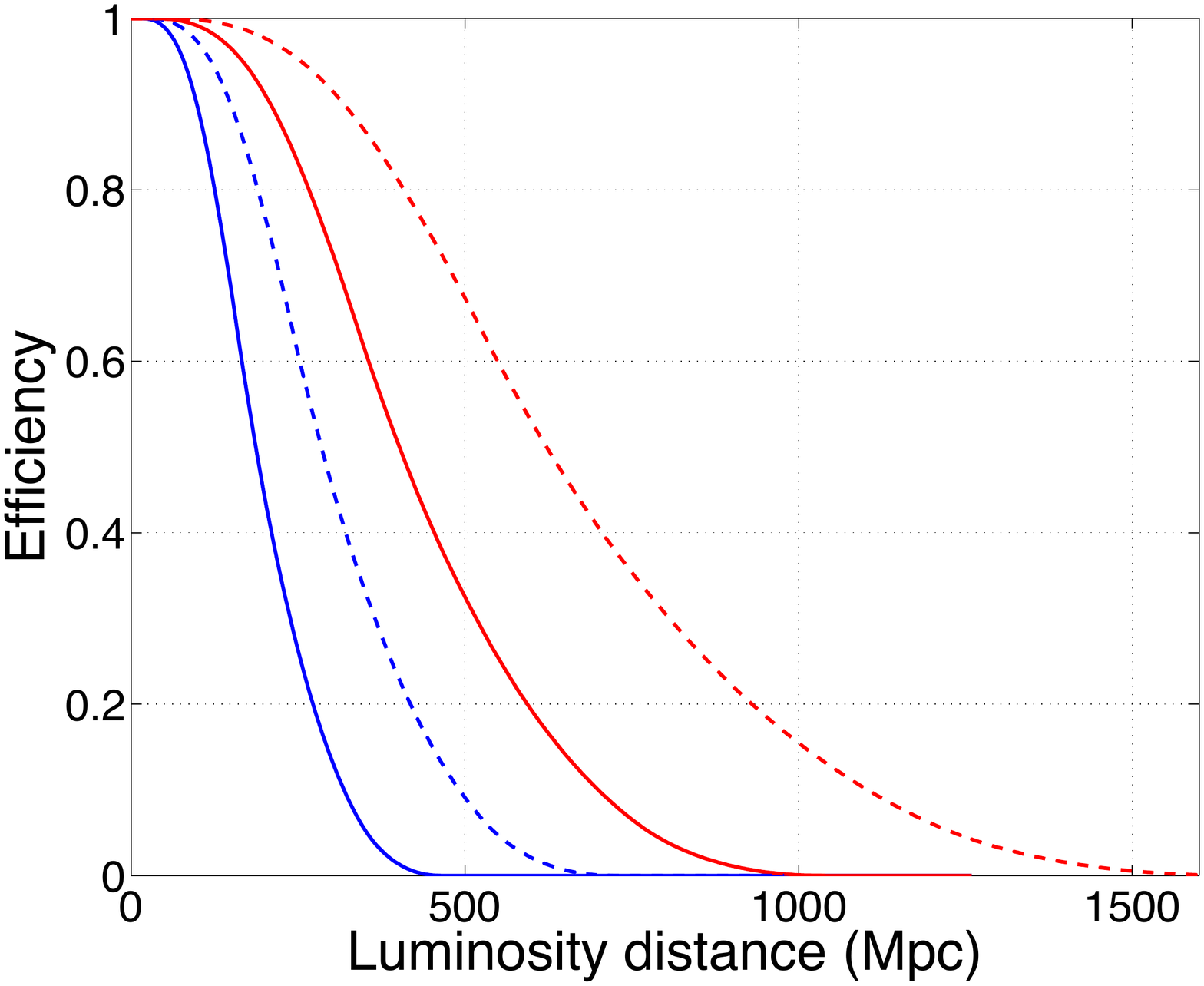}
\includegraphics[width=0.49\textwidth]{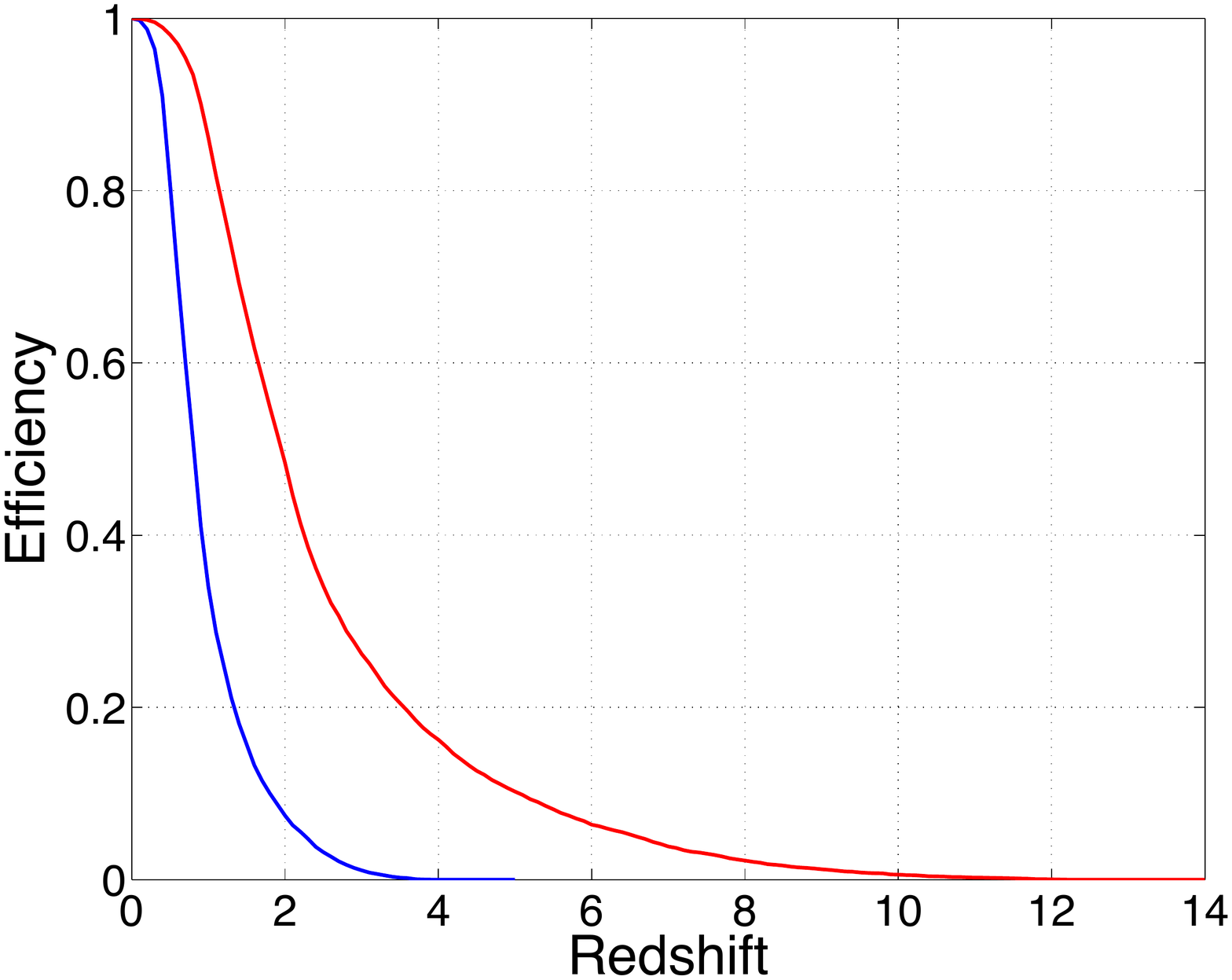}
\caption{Left: GW detection efficiency as a function of luminosity distance of BNS (blue) and NS-BH (red) for the ALV network. The continuous and dashed lines correspond to signal-to-noise ratio threshold of 12 and 8 respectively. 
Right: GW detection efficiency as a function of redshift of BNS (blue) and NS-BH (red) for ET and a signal-to-noise ratio threshold of 8. We assumed masses of 1.4 M$_\odot$ for neutron stars and 10 M$_\odot$ for black holes.}
\label{fig:DetEff_GW}
\end{figure}

Several factors affect the maximum distance to which a GW detector will be able to detect a source (the horizon). 
\begin{itemize}
\item The relative position of the source with respect to the detector plane at the time of coalescence:
a detector is most sensitive to a GW signal that propagates orthogonally to the plane of the detector, meaning that the signal will be affected by both the position in the sky and the time of arrival. 
\item The second factor is the inclination angle, i.e. the angle between the normal of the source orbital plane and the observer's line of sight.
\end{itemize}
Fig.~\ref{fig:DetEff_GW} displays the efficiency of ALV and ET for the two cases BNS and NS-BH. The SNR threshold is set to $\rho_T=12$ for ALV \citep{Kelley13}, corresponding roughly to an SNR of 8 on at least two detectors. The observation of an EM counterpart, by increasing detection confidence, may allow to reduce the SNR threshold, thus we also use a less conservative value of $\rho_T=8$, corresponding to an SNR of about 6 on two detectors (Sutton P., private communication). For the planed ET, we use $\rho_T=8$ as suggested by mock data challenges \citep{Regimbau12}.
At a redshift $z \sim 0$, all the sources are detected, but as the distance increases, only the best located and oriented sources reach the required threshold. 
The horizon of the detector corresponds to the maximum distance of detection for the optimally oriented (face-on) and positioned sources, and increases as the SNR threshold decreases. The ALV horizon is about 460 Mpc for BNS and 1 Gpc for NS-BH, assuming $\rho_T=12$, or 720 Mpc for BNS and 1.6 Gpc Mpc for NS-BH, assuming $\rho_T=8$. The ET horizon is $z=4$ for BNS and $z=13.5$ for NS-BH.{\footnote {our horizon distance is a bit higher than the one obtained in \citet{Abadie10a}, as we have taken into account the redshift (see Eq.~\ref{eq:SNR}). For a single aLIGO detector, we would obtain an horizon of 485 Mpc instead of 445 Mpc for BNS, and 1090 Mpc instead of 927 Mpc for NS-BH.} }


\subsection{Coincidence efficiency: the case of a perfect GRB detector}
\label{IIIB-perfect}

\begin{figure}
  \includegraphics[width=0.5\columnwidth]{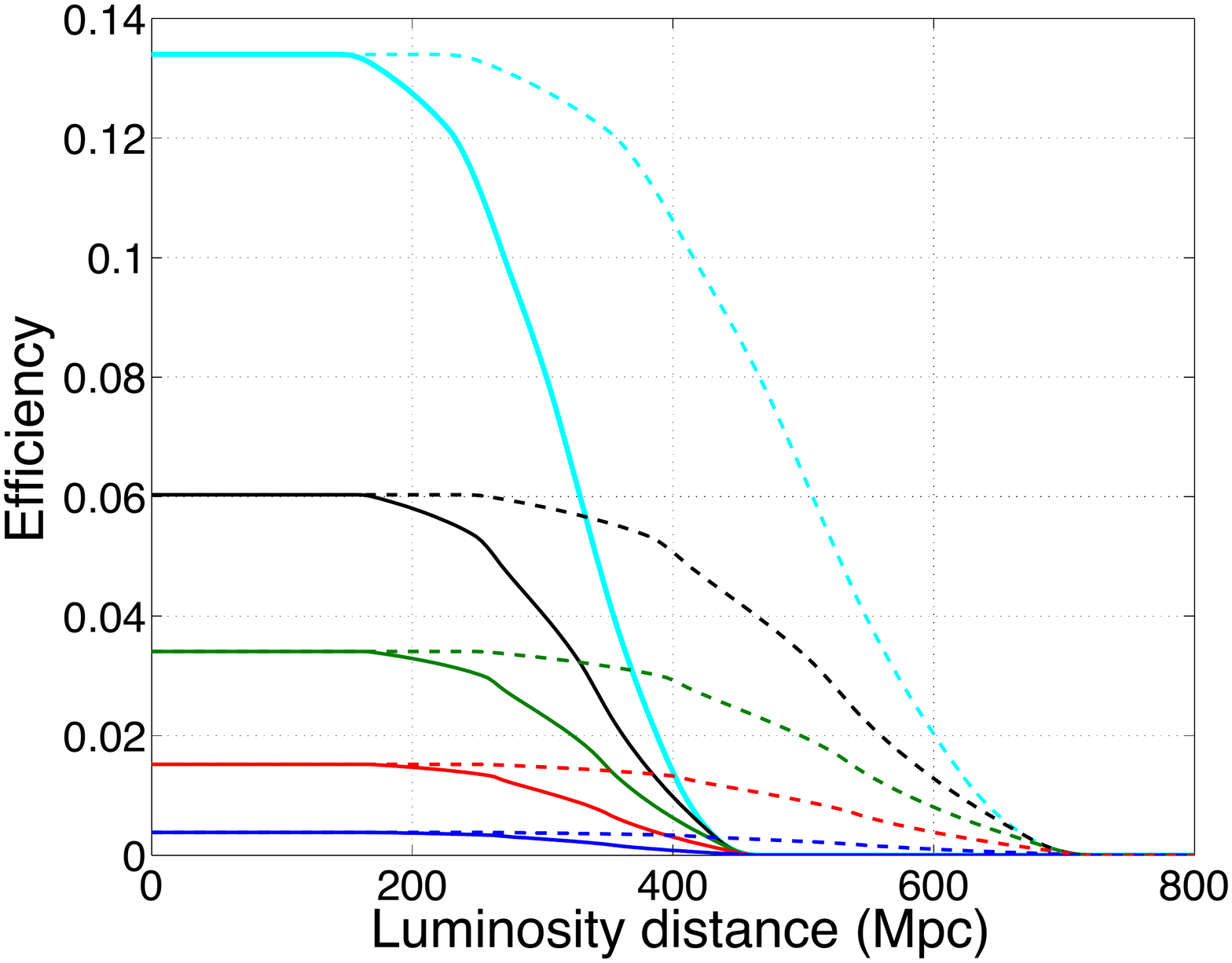}
  \includegraphics[width=0.5\columnwidth]{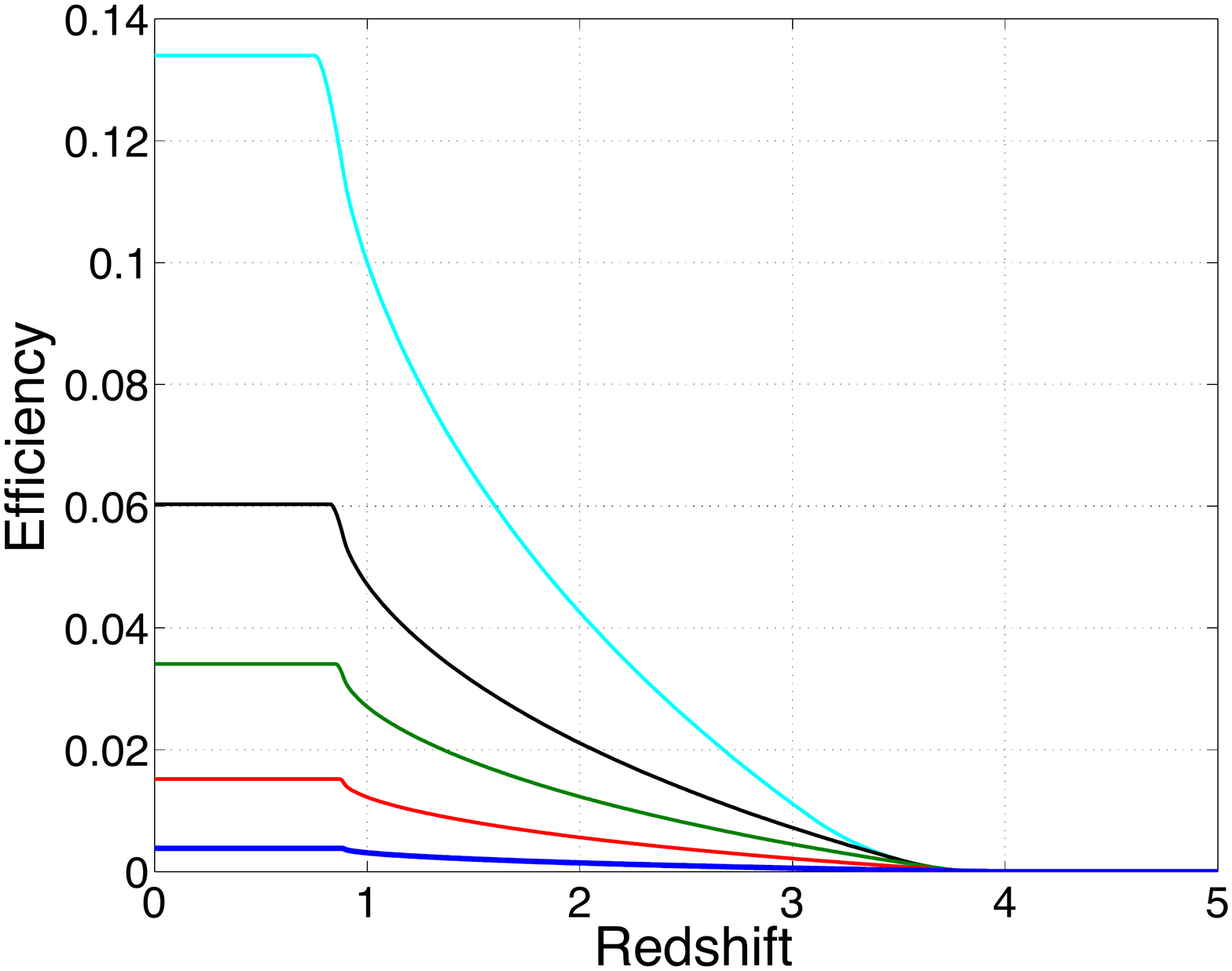}\\
  \includegraphics[width=0.5\columnwidth]{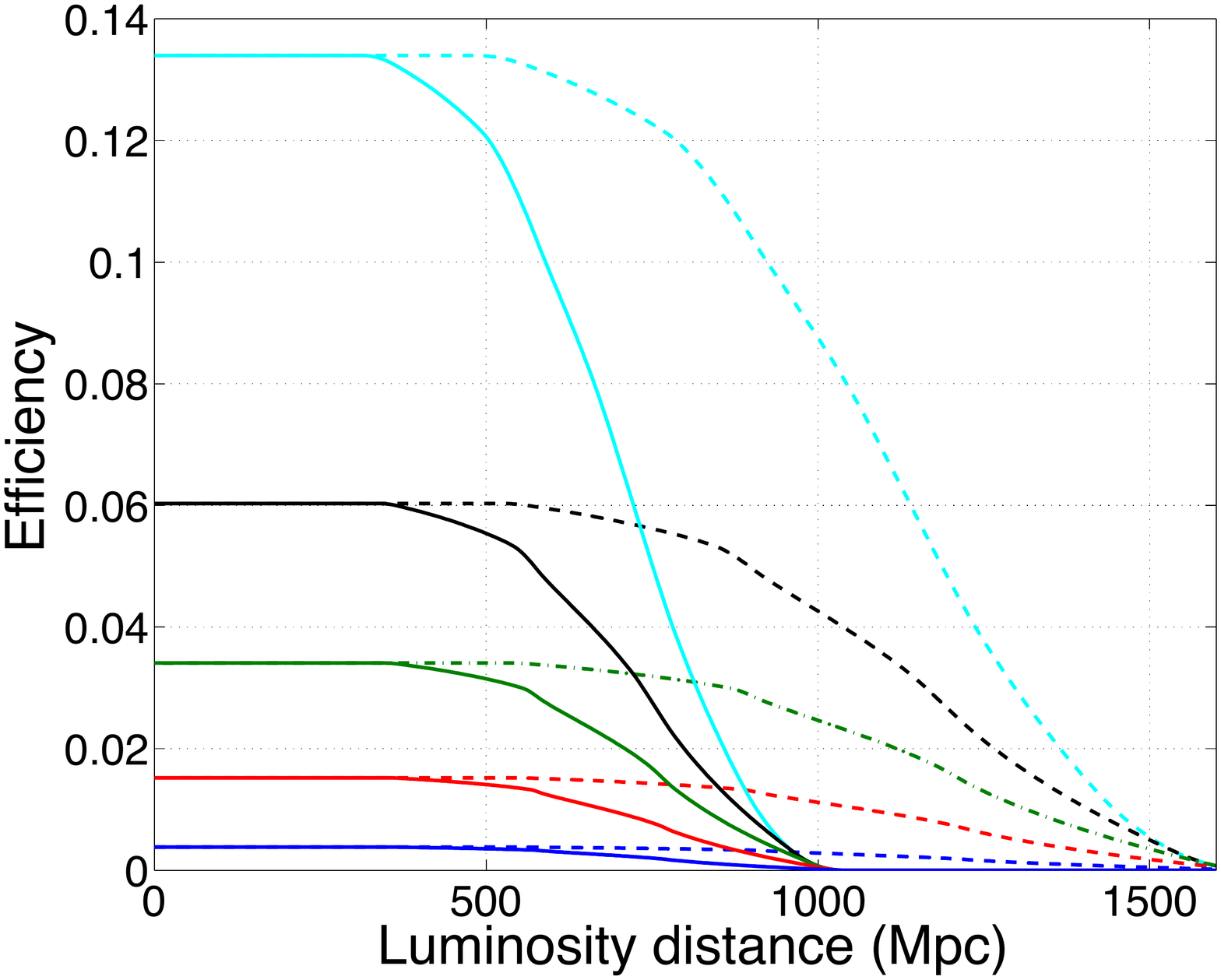}
  \includegraphics[width=0.5\columnwidth]{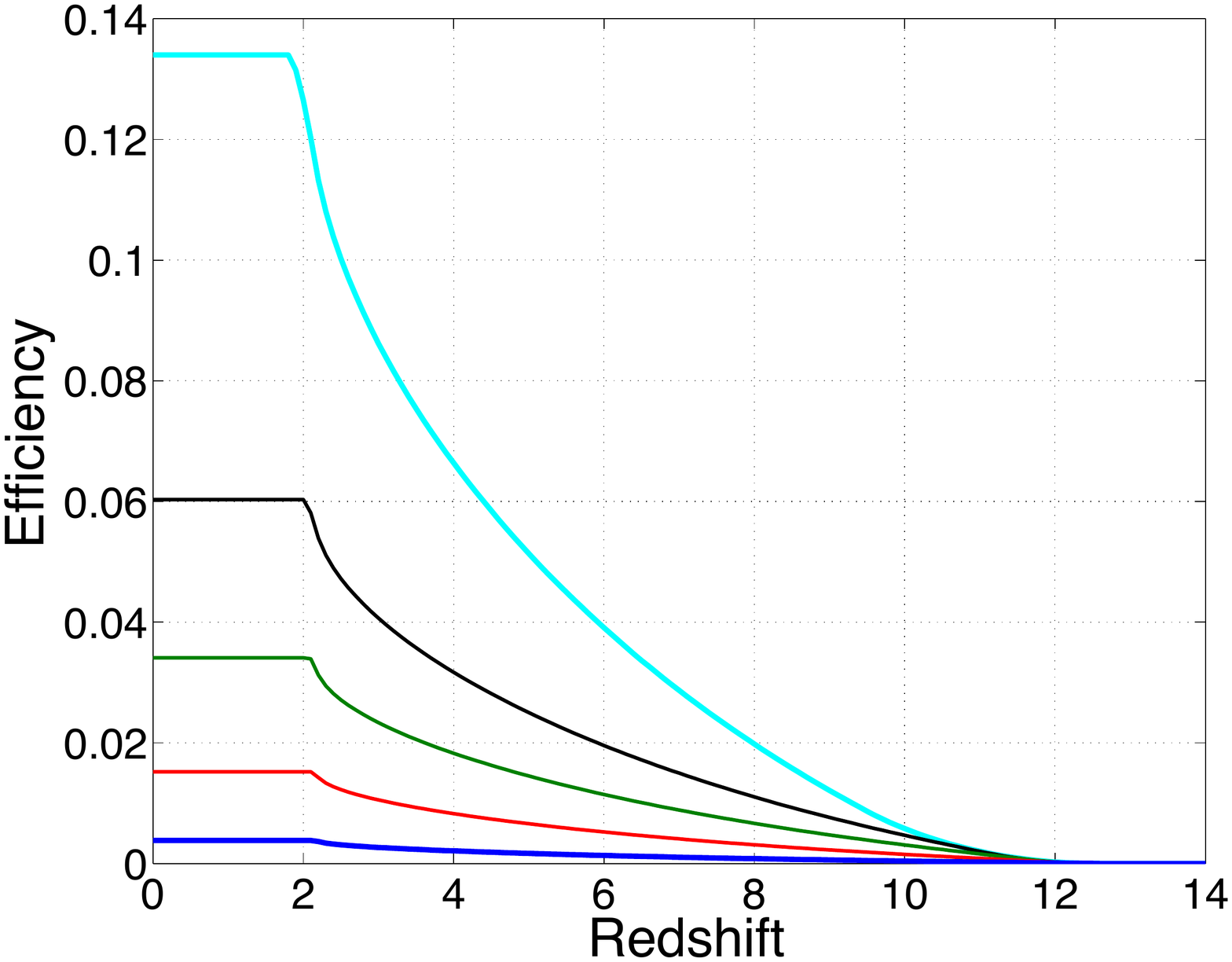}
  \caption{Left: GW/GRB coincident detection efficiency of BNS (top) and NS-BH (bottom), for ALV, assuming infinite sensitivity, an FOV of $4 \pi$ and a duty cycle of 100\% for the GRB satellite, and signal-to-noise ratio threshold of 12 (continuous lines) and 8 (dashed  lines). The curves that extend to larger distances are for a threshold of 8.
                Right: GW/GRB coincident detection efficiency as a function of redshift of BNS (top) and NS-BH (bottom), for ET, infinite sensitivity, an FOV of $4 \pi$ and a duty cycle of 100\% for the GRB satellite, and signal-to-noise ratio threshold of 8.
                The different lines indicate different values of the beaming angle. From top to bottom, 30, 20, 15, 10, and 5 degrees.}
  \label{fig:Eff_ideal}
\end{figure}

In order to model the EM selection effects, we first consider the case of a perfect GRB detector with FOV of $4\pi$ sr, duty cycle of $100\%$, and infinite sensitivity, so that the only selection effect is the beaming fraction. 
Compared to the previous case, we also require that the inclination $\iota$ ( or $\pi- \iota$) is equal or smaller than the beaming angle, i.e. one of the two opposite jets is directed towards the Earth. 
The resulting efficiency presented in Fig.~\ref{fig:Eff_ideal}, $ \varepsilon_{cdp}(z)$, is equal to the beaming factor at $z=0$ and shows a {\it plateau} until a redshift $z_*$. Short GRBs closer than $z_*$ can all be detected in GWs, then the efficiency decreases steadily until the GW detector horizon. The end of the {\it plateau} correspond to the redshift that gives a SNR $\rho = \rho_T$ when the inclination is equal to the beaming angle and the position in the sky and the polarization are such that ${\cal F}$ is minimal.

\begin{figure}
 \includegraphics[width=0.5\columnwidth]{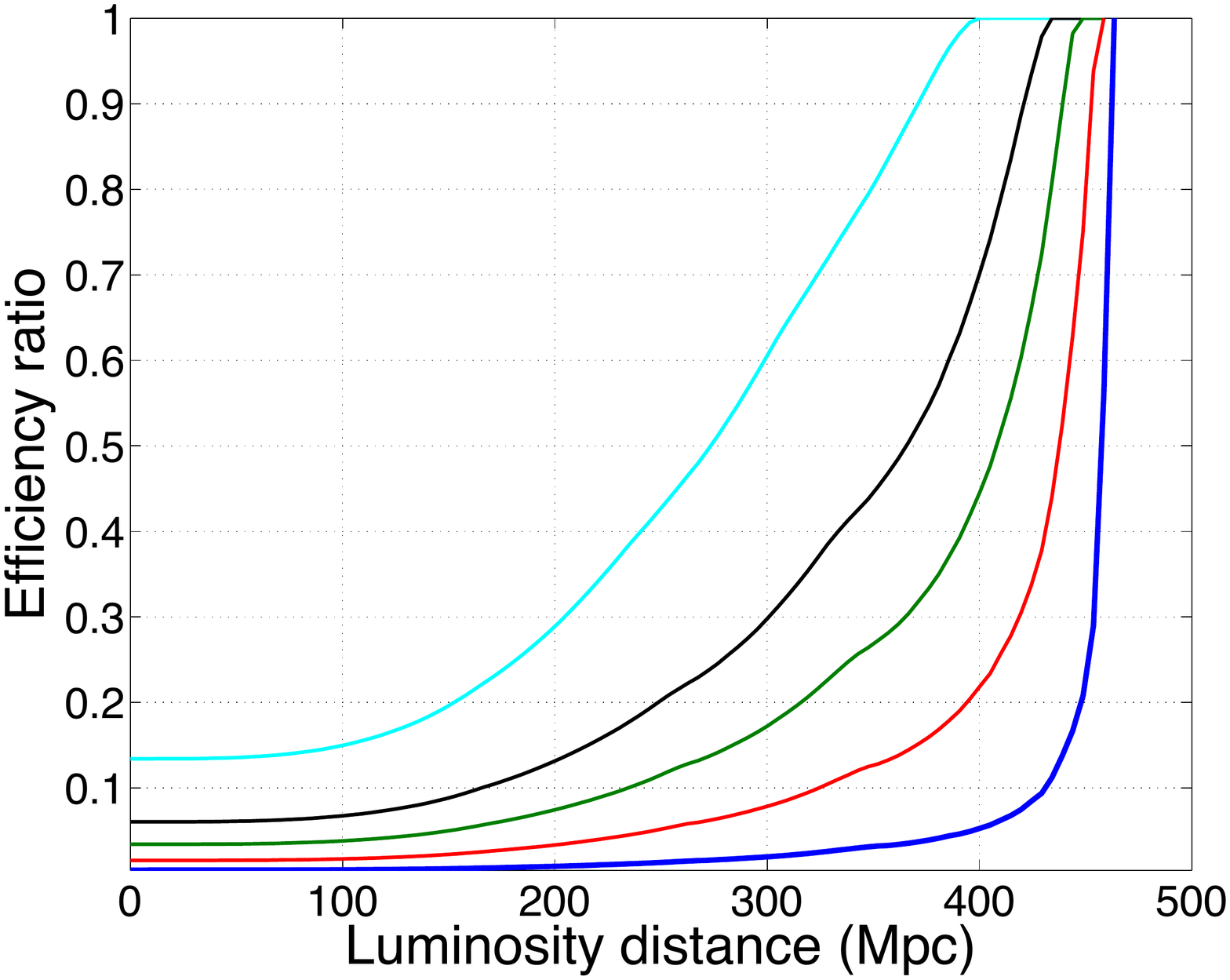}
 \includegraphics[width=0.5\columnwidth]{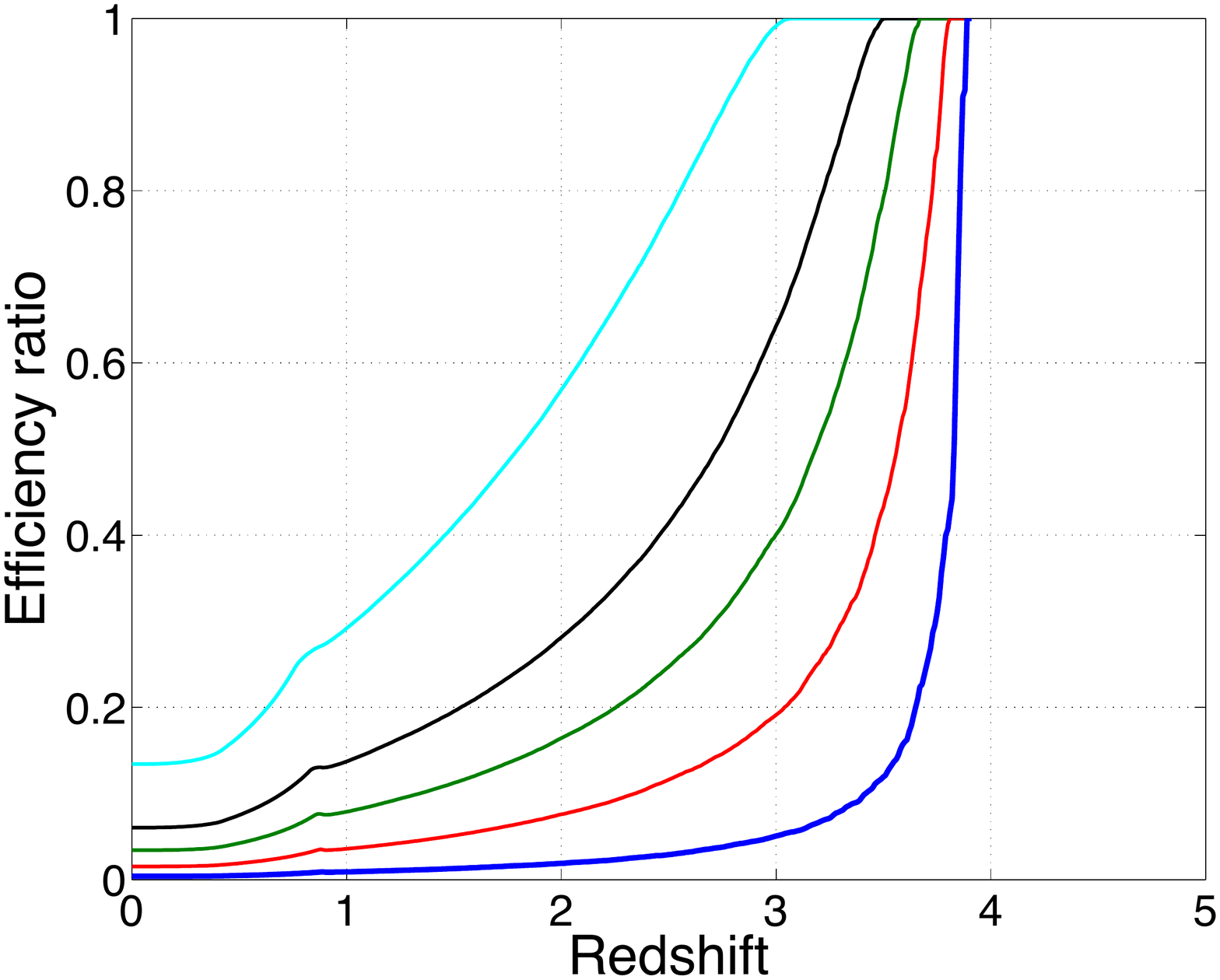}
  \caption{Ratio of total GW events to those that can be observed as sGRBs, assuming infinite sensitivity, an FOV of $4 \pi$ and a duty cycle of 100\% for the GRB. Left: ALV with signal-to-noise ratio threshold of 12. Right: ET with signal-to-noise ratio threshold of 8. The behavior is similar for NS-BH.             
 The different lines indicate different values of the beaming angle. From top to bottom, 30, 20, 15, 10, and 5 degrees.}
  \label{fig:ratio}
\end{figure}

Fig.~\ref{fig:ratio} shows the efficiency ratio between $\varepsilon_{GW}(z)$ and $\varepsilon_{cdp}(z)$. The fraction of GW/EM coincident detections corresponds to the beaming angle efficiency at $z = 0$, then increases with the distance to reach 1 close to the GW horizon, where the only sources that can be detected are the best oriented and satisfy Eq.~\ref{eq-SE_angle}. 

Our assumption of the same averaged value of the beaming angle for the whole population of sGRBs is certainly not realistic. We considered also a set of simple distributions (uniform and Gaussian for $\theta_B$ and $\cos \theta_B$ respectively) as well as the log-Gaussian distribution of \citet{Goldstein11}, derived from observations, with average value $\mu_{\log \theta_B}=2.0794$ and standard deviation $\sigma_{\log \theta_B}=0.69$ ($\theta_B$ in degrees).
We found that the efficiency is essentially sensitive to the average value of the beaming factor $<\Theta_B>=(1-<\cos \theta_B>)$ and that the width and the shape of the distribution has very little  impact. 
The efficiency is equal to $<\Theta_B>$ during the {\it plateau} and the only difference is visible at the very end, where the distribution of the beaming angle results on a smoother transition between the {\it plateau} and the sharp decrease of the efficiency curve. 

On the other hand, NS are expected to have a very narrow mass distribution centered around 1.4 M$_{\odot}$ but the BH mass distribution is more uncertain. Changing the mass of the system affects the redshift $z_*$ at which the {\it plateau} ends as well as the horizon distance. For a delta function, the SNR is shifted toward lower values when the mass decreases, as well as $z_*$ and the horizon. 
For a broader distribution (for exemple a Gaussian or a uniform distribution), there is a smooth transition between the {\it plateau} and the sharp decrease of the efficiency curve, starting at the critical redshift $z_*(\mathcal{M}_{\min})$ corresponding to the minimal value of the chirp mass $\mathcal{M}_{\min}$ and ending at $z_*(\mathcal{M}_{\max})$ corresponding to the maximal value of $\mathcal{M}_{\max}$. 
The horizon is the maximal distance observed for $\mathcal{M}_{\max}$.

\subsection{Case of a realistic GRB detector}\label{IIIB-realistic}

We now consider the case of a realistic GRB detector with finite sensitivity, and reduced FOV and duty cycle.
The FOV, the duty cycle and the flux limit of the GRB satellites being independent of the GW detection, we write the final efficiency as the product:
\begin{equation}
\varepsilon_{cd}(z) = \varepsilon_{FOV} \times \varepsilon_{DC} \times \varepsilon_{cdp}(z) \times \varepsilon_{sat}(z) ,
\end{equation}
\noindent where $\varepsilon_{FOV}$ is the FOV divided by $4 \pi$ sr, $\varepsilon_{DC}$ is the duty cycle, $\varepsilon_{cdp}$ is the efficiency for a perfect detector found in Section \ref{IIIB-perfect}, and $\varepsilon_{sat}(z)$ is the fraction of sGRBs whose flux  is larger that the limiting flux $F_{\mathrm{lim}}$ (see Section \ref{GRB_selection}). 

\begin{figure}
\includegraphics[width=\columnwidth]{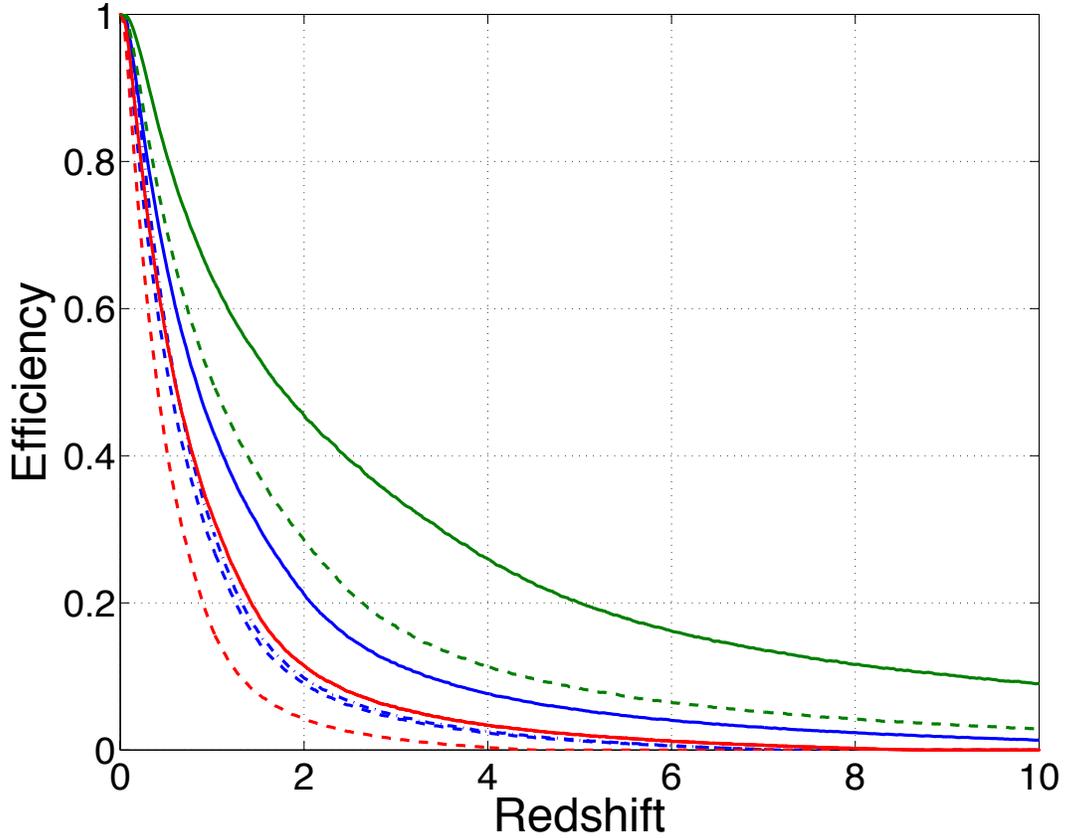}
\caption{Detection efficiency of the \textit{Swift} satellite for sGRBs assuming a flux limit of 1.5 ph s$^{-1}$cm$^{-2}$ (blue lines), a pessimistic value of 2.5 ph s$^{-1}$cm$^{-2}$ (red lines), corresponding to sGRBs with redshift measurement, and an optimal value of 0.56 ph s$^{-1}$cm$^{-2}$ (green lines), corresponding to on-axis sources. The continuous lines correspond to a peak luminosity probability distribution with $L_*=10^{51}$ erg/s and $\Delta_1=100$), and the dashed line to $L_*=5 \times 10^{50}$ and $\Delta_1=100$. For comparison, we have also indicated the efficiency for a larger value of the low luminosity bound ($L_*=5 \times 10^{50}$ and $\Delta_1=30$) in dash-dotted blue. The efficiency is calculated for an FOV of $4 \pi$ and a duty cycle of 100\%, in order to have an efficiency of 1 at $z=0$.}
\label{fig:DetEff_Swift}
\end{figure}

The efficiency of {\it Swift} derived from our Monte Carlo procedure is presented in Fig.~\ref{fig:DetEff_Swift} for a flux threshold of 1.5 ph~s$^{-1}$~cm$^{-2}$, a pessimistic value of 2.5 ph~s$^{-1}$~cm$^{-2}$, corresponding to sGRBs with redshift measurement, and an optimal value of 0.56 ph~s$^{-1}$~cm$^{-2}$, corresponding to on-axis sources. 
The shape is very similar to that presented in \citet{Howell14} and \cite{Lien14} for long GRBs, with a sharp exponential decrease. 

\begin{figure}
\includegraphics[width=0.5\columnwidth]{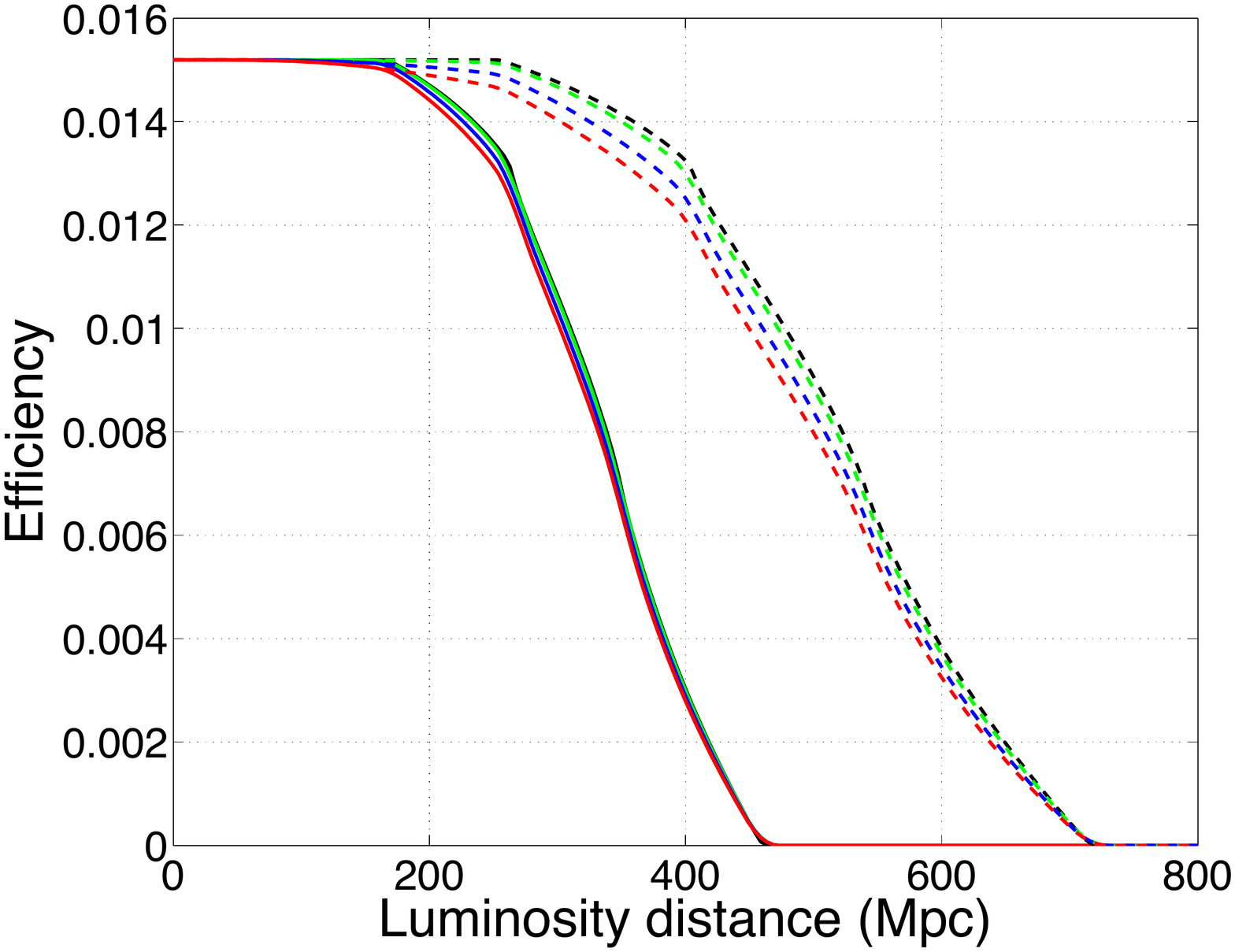} 
 \includegraphics[width=0.5\columnwidth]{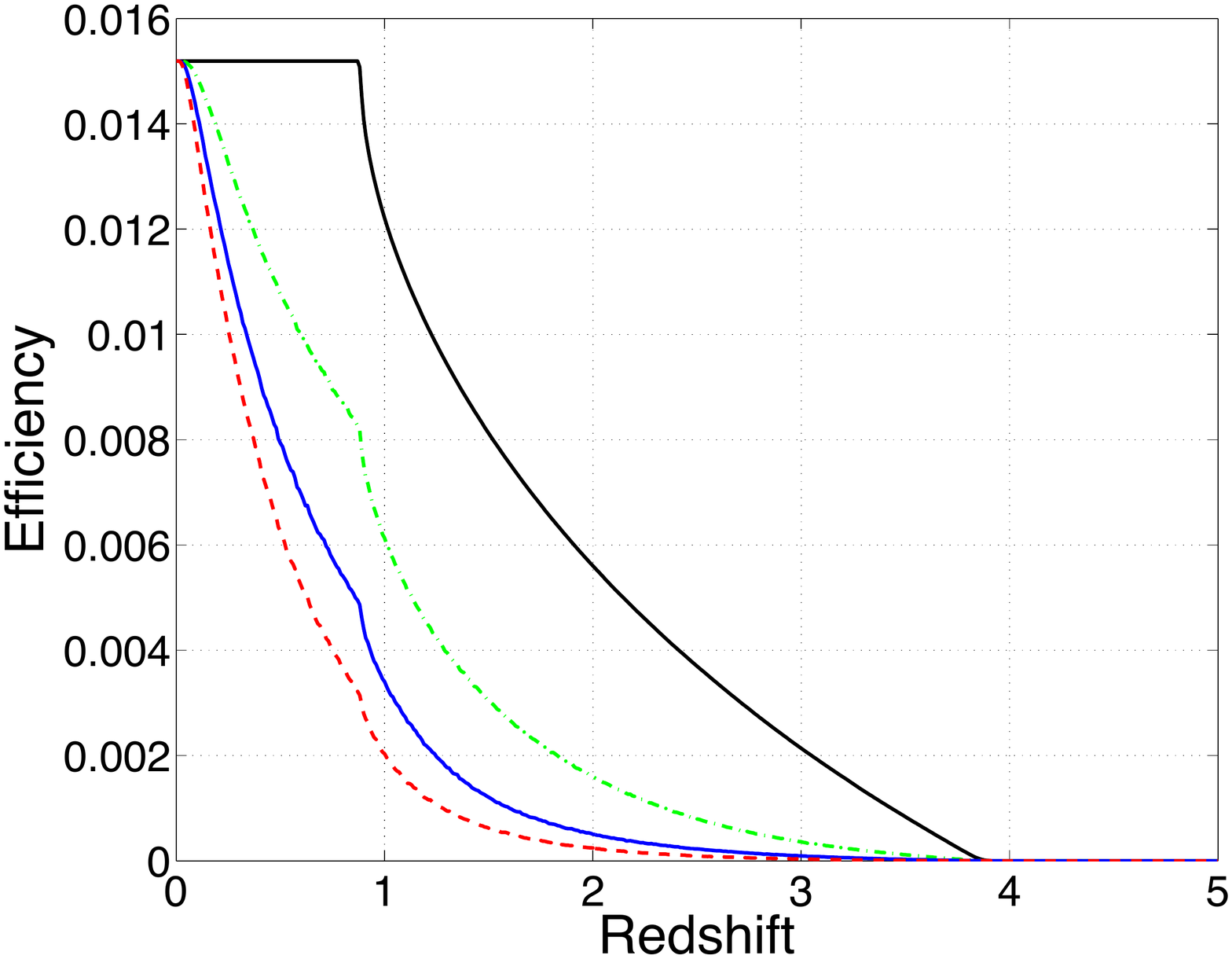}
  \includegraphics[width=0.5\columnwidth]{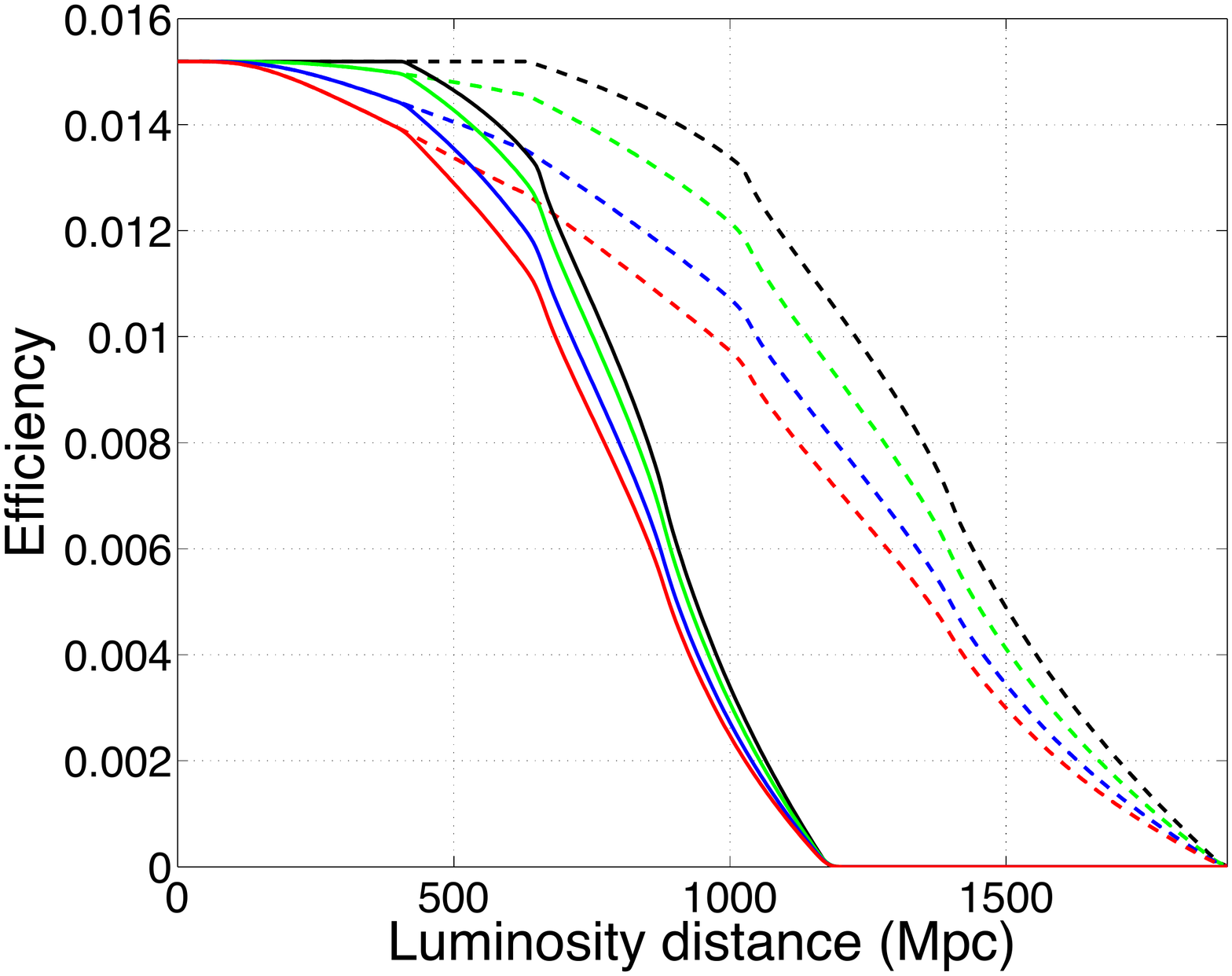} 
 \includegraphics[width=0.5\columnwidth]{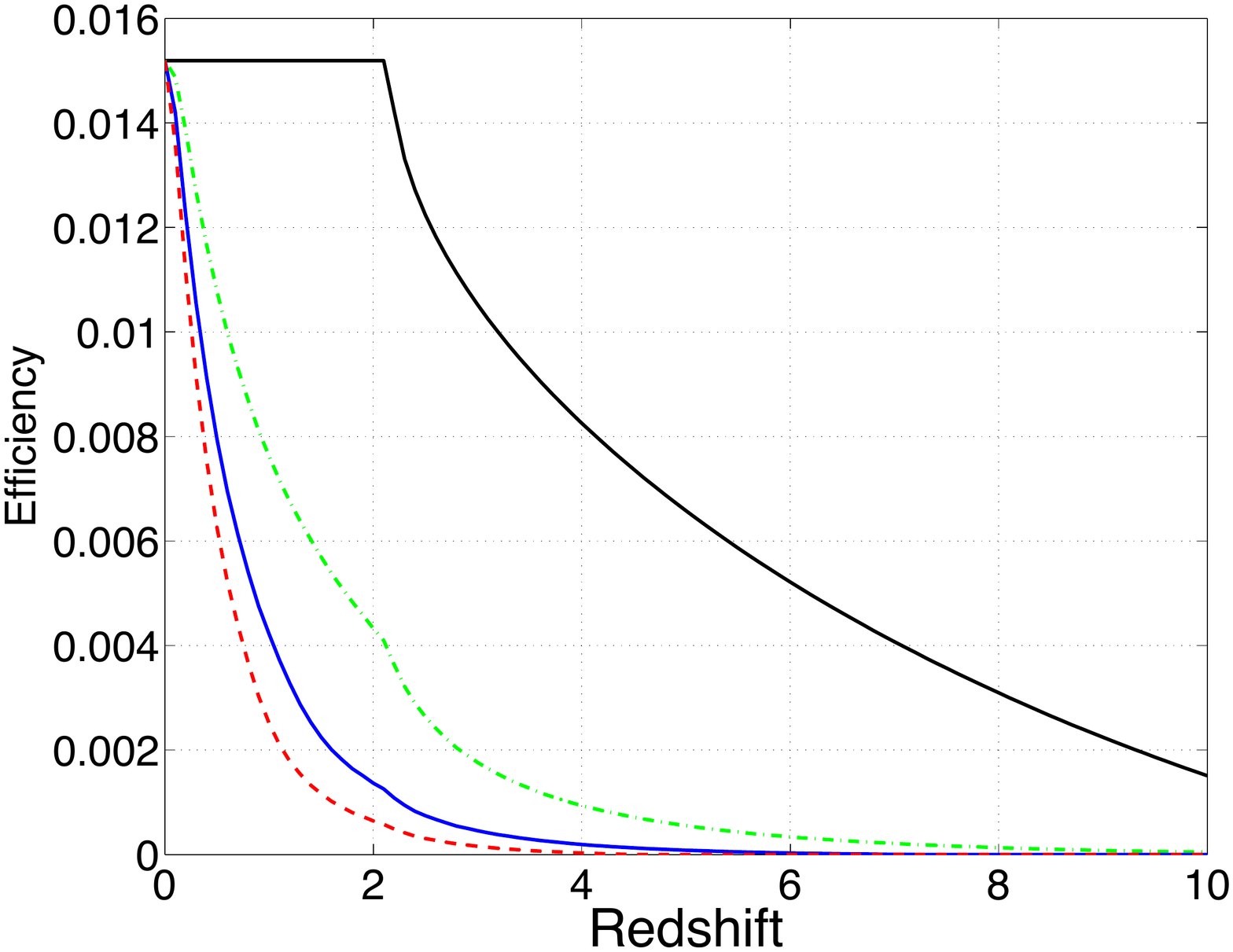}
  \caption{GW/GRB detection efficiency of BNS (left) and NSBH (right), for ALV (top) and ET (bottom), and the \textit{Swift} satellite sensitivity with a flux limit of 1.5 ph s$^{-1}$cm$^{-2}$ (continuous blue line) a pessimistic value of 2.5~ph s$^{-1}$cm$^{-2}$ (dashed red line), corresponding to sGRBs with redshift measurement, and an optimal value of 0.56 ph s$^{-1}$cm$^{-2}$ (dash-dotted green line), corresponding to on-axis sources. The black curve corresponds to the efficiency for an infinite sensitivity satellite and is shown for comparison. The efficiency is calculated for an FOV of $4 \pi$ and a duty cycle of 100\%.}
  \label{fig:Eff_flux}
\end{figure}

Fig.~\ref{fig:Eff_flux} shows the coincident efficiency, taking into account the sensitivity of {\it Swift}, for $\theta_B=10^{\circ}$ and the three flux threshold of 0.56, 1.5 and 2.5 ph s$^{-1}$cm$^{-2}$. 

For ALV, the coincident sensitivity is limited rather by the GW detector horizon than the GRB flux threshold. The effect of the {\it Swift} sensitivity is not significant for BNS, but is noticeable for NS-BH, especially with an SNR threshold of 8, mainly by reducing the size of the {\it plateau}. 

For ET, the efficiency of a GRB satellite like {\it Swift} will drop much faster than the GW efficiency, causing the suppression of the {\it plateau} and shifting the horizon to smaller redshift. 
In order to have 100\% of the sources above the threshold at $z ~\sim 1$ (corresponding to the end of the {\it plateau} for BNS), and then fully exploit the potentiality of ET, $F_{\mathrm{lim}}$ would have to be reduced to$\sim 0.0013$ ph~s$^{-1}$cm$^{-2}$ over the next decade if $L_*=10^{51}$ erg s$^{-1}$ (a factor of 2 smaller if $L_*=5 \times 10^{50}$ erg s$^{-1}$). A value of 0.2 ph~s$^{-1}$cm$^{-2}$ would give 80\% of the sources above the threshold at $z=1$ and 1.1 ph~s$^{-1}$cm$^{-2}$ 50 \%.





\section{Rate}\label{sec:rates}

In this paper, we assume the coalescence occurs after two massive stars in a binary system have burned all their nuclear fuel, have evolved into red giants, and the cores have collapsed, possibly after supernova explosions, forming a bound system of two compact objects (neutron stars or black holes) inspiralling each other due to the emission of GWs. Another scenario suggests that NS or BH binaries could form through dynamical captures in dense stellar environment. However, most simulations indicate that the chance for this to occur is small, due to the presence of massive black holes at the center that substitute into binaries during dynamical interactions, so that this population may not represent a significant fraction of the total coalescence rate \citep{Abadie10a}.\footnote{Notice that the dynamical origin is favored in the recent study of \citet{Wanderman14} as sGRB progenitors, which may suggest that they are more numerous that what is predicted by simulations or that dynamical binaries are not affected by the same selection effects as primordial binaries \citep{Grindley06}}

The coalescence rate per interval of redshift:
\begin{equation}
\frac{dR}{dz}(z) = \dot{\rho}(z) \frac{dV}{dz} ,
\end{equation}
is obtained by multiplying the element of comoving volume $\frac{dV}{dz}$ and the coalescence rate per unit of volume: 
\begin{equation}
\dot{\rho}(z) \propto \int \frac{\dot{\rho}_*(z_f)}{1+z_f}P(t_d)dt_d \,\ \mathrm{with}\,\ \dot{\rho}(0)=\dot{\rho}_0 ,
\label{eq:dRdz}
\end{equation}
\noindent In this equation, $\dot{\rho}_*$ is the star formation rate (SFR), $\dot{\rho}_0$ the local coalescence rate in Mpc$^{-3}$ Myr$^{-1}$, $z_f$ the redshift at the time of formation of the massive binary system and $P(t_d)$ the probability distribution of the delay between the formation and the coalescence. The number of massive systems that remain bounded after two supernovas (or prompt core-collapses) is uncertain, as well as the time to coalescence (the delay) which depends on complicated evolution scenario involving common envelope and mass transfer. We assume a distribution of the form $P(t_d) \propto 1/t_d$ with a minimal delay of 20 Myr for the population of BNS and 100 Myr for BH-NS, as suggested by the population synthesis software StarTrack \citep{Dominik12}, and we leave $\dot{\rho}_0$  which is given between $0.001-10$ Mpc$^{-3}$ Myr$^{-1}$ by \citet{Abadie10a} as a free parameter. 
The co-moving volume element is given by:
\begin{equation}
\frac{dV}{dz}(z)=4 \pi \frac{c}{H_0} \frac{r(z)^2}{E(\Omega,z)}\,,
\end{equation}
where
\begin{equation}
r(z)= \frac{c}{H_0}\int_0^z \frac{dz'}{E(\Omega, z')}\,,
\label{eq:distance}
\end{equation}
and 
\begin{equation}
E(\Omega,z)=\sqrt{\Omega_{\Lambda}+\Omega_{m}(1+z)^3}\,.
\end{equation}
In this paper, we use a standard flat cold dark matter ($\Lambda$CDM) model for the Universe, with $\Omega_m = 0.3$ and $H_0 = 70$ km s$^{-1}$ Mpc$^{-1}$.

\begin{figure}
\centering
\includegraphics[angle=0,width=\columnwidth]{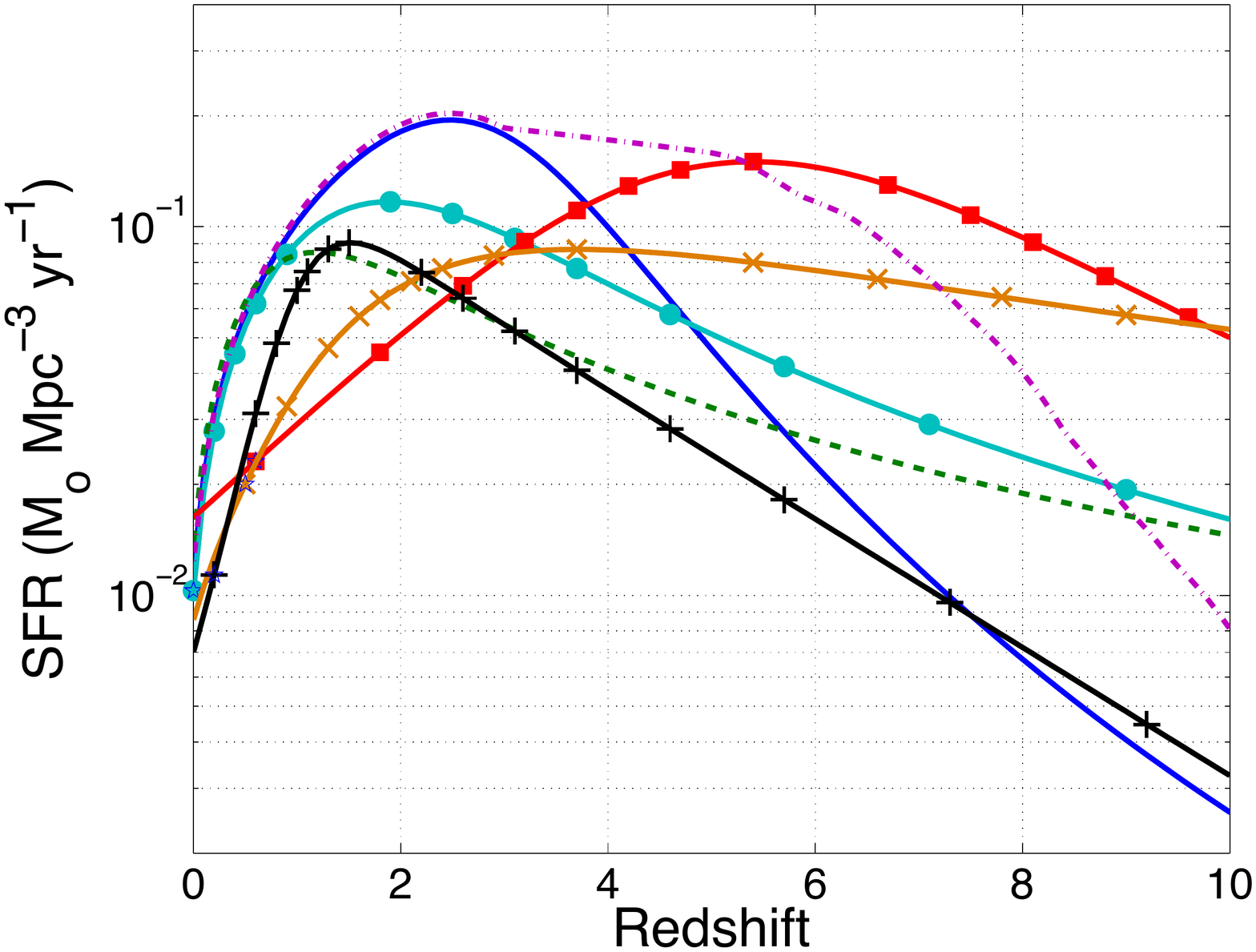}
\caption{Cosmic star formation rates (in M$_\odot$ Mpc$^{-3}$
yr$^{-1}$) used in this paper :  SFR of \citet{Hopkins06} (our reference model) in continuous blue, SFR of \citet{Fardal07} with light blue dots, SFR of \citet{Wilkins08}  in dashed green,  SFR of \citet{Springel03} with red squares, SFR of \citet{Nagamine06} with orange crosses, SFR of \citet{Tornatore07} in dot-dashed purple, and the SFR of \citet{Madau98} in black with plus signs. 
\label{fig-allsfr}}
\end{figure}

In order to account for the uncertainty in the star formation history, we consider seven different SFRs described in detail in \citet{Regimbau09,Regimbau11} and plotted in Figure \ref{fig-allsfr}.
As a reference, we use the SFR of \cite{Hopkins06}, which is derived from measurements of the galaxy luminosity function in the ultra-violet (UV) and far infra-red (FIR) wavelengths, and is normalized by the Super Kamiokande limit on the electron antineutrino flux from past core-collapse supernovas.  This
model is expected to be quite accurate up to $z \sim 2$, with very tight constraints at redshifts $z<1$ (to within $30-50 \%$).  
\citet{Fardal07}) use a different set of measurements and a different dust extinction correction. The SFR found in \citet{Fardal07} is the same as that of \citet{Hopkins06} up to $z \sim 1$, but decreases slightly at higher redshifts. We also consider the model described by \citet{Wilkins08}, which is derived from measurements of the stellar
mass density. The SFR is equivalent to that in \cite{Hopkins06,Fardal07} for $z \lesssim 0.7$, but again is lower at higher redshifts. 
Note that at present there is a discrepancy between the ``instantaneous'' SFR, measured from the emission of young stars in star forming regions, and the SFR as determined from extragalactic background light.  This could have an important impact on the contribution of at $z > 2$.
Finally, we consider the analytical SFR of \citet{Springel03} derived from cosmological Smoothed Particle Hydrodynamics numerical simulations, the model of \citet{Nagamine06} derived from the fossil record of star formation in nearby galaxies which probably underestimate the SFR at small redshifts but may be more accurate that and is constant at high redshifts due to the contribution of elliptical galaxies, and the SFR of \citet{Tornatore07}, which combined observations and simulations at higher redshift.  
For completeness, we also considered a previous model derived from the UV continuum and H$\alpha$, up to $z \sim 4$, where the main uncertainty comes from dust extinction, which spreads the UV luminosity into the FIR \citep{Madau98}.

\subsection{Detection rates}

The GW detection rate is obtained by integrating over redshift the product of the coalescence rate given in Eq.~\ref{eq:dRdz} and the GW efficiency:
\begin{equation}
R_{\mathrm{gw}} = \int_0^{z_{\max}} \varepsilon_{\mathrm{gw}}(z)  \frac{dR}{dz}(z) dz ,
\label{eq:Rgw}
\end{equation}
where $z_{\max}$ corresponds to the beginning of stellar activity. From our models, $z_{\max} \sim 10-20$.

This equation is different than the approximated expression $R_{\mathrm{gw}} = \frac{4}{3} \pi (d_{\max}/2.26)^3$ \citep{Fin93}, where $d _{\max}$ is the horizon distance and the factor 2.26 a correction needed to average over sky location and orientation. 
These two equations are similar when facing with the probe of a small volume, like in the current era of advanced detectors, but strongly diverge for a larger horizon, which is the key signature of ET. 
In such a case, including the efficiency, the star formation history and the cosmology becomes crucial.

 
In the same way, we calculate the GW/sGRB detection rate as:
\begin{equation}
R_{cd} = \int_0^{z_{\max}}  \varepsilon_{\mathrm{cd}}(z)  \frac{dR}{dz}(z) dz .
\label{eq:rate}
\end{equation}
Compared to the simple beaming factor correction $R_{cd} = \Theta_B R_{\mathrm{gw}}$, this equation results in an improvement of the final coincident rate by a factor of $\sim 3$ in the case of a perfect sGRB detector, only limited by the beaming selection effect.

The GW coincident rates for ALV and the {\it Swift} detector are presented in Table~\ref{tab:rates}, for different beaming angles. We used as a reference, the local rates of $\dot{\rho}_c^o=0.06$ Mpc$^{-3}$ Myr$^{-1}$ for BNS and 0.003 Mpc$^{-3}$ Myr$^{-1}$ for NS-BH, obtained recently by \citet{Dominik14} for the StarTrack standard model. They correspond to GW detection rates of about 3 yr$^{-1}$ for BNS and 2 yr$^{-1}$ for NS-BH, assuming a threshold of 12 for the SNR. That turns out to represent a coincident detection rate for BNS and NS-BH mergers from a few every 1000 years to a few every 10 years depending on the beaming angle $\theta_B$. These rates are 2\% ($\rho_T=12$) and 5\% ($\rho_T=8$) smaller than the rates one would obtained assuming infinite GRB satellite flux sensitivity for BNS, and 10\% ($\rho_T=12$) and 18\% ($\rho_T=8$) for BH-NS. Decreasing the SNR threshold from 12 to 8, increases the number of detections by a factor of 3. The coincident rates derived for {\it Swift} would improve by a factor of $\sim 10$ for a detector with a large FOV like Fermi, but only by a factor of 1.4 and 2.4 for the planned FOV of SVOM and LOFT. 


Our results derived from simulations of the cosmological population of BNS and NS-BH are in agreement with the rates derived from sGRBs observations by \citet[][$0.06-0.16$ yr$^{-1}$]{Siellez14}, for our reference value of the local merger rate of $\dot{\rho}_c^o \sim 0.06$ Mpc$^{-3}$ Myr$^{-1}$, or the lower bound of \citet{Coward12} ($\sim 0.1-10$ yr$^{-1}$ after FOV and DC correction), based on bias correction of {\it Swift} data.  They are a factor of 10 smaller than the rates found by \citet[][$0.2-1$ yr$^{-1}$]{Petrillo13}. Comparing the different studies is difficult though, as different authors used different assumptions. 

As one can see, the coincident rate is very sensitive to the beaming angle $\theta_B$. Between a half jet opening angle of $\theta_B$=5$^\circ$ and $\theta_B$=30$^\circ$, the rate increases by a factor of about $\sim 35$. 
Using the efficiency presented in Fig.\ref{fig:Eff_flux}, we roughly estimated that for our reference value of the local merger rate, the beaming angle, should be between $3^{\circ}-10^{\circ}$  in order to reproduce the actual observed rate of 8 sGRBs per year with {\it Swift}, which is consistent with current models of the sGRB jet. This would favor a coincident rate between ALV and {\it Swift} $<0.1$ yr$^{-1}$. Increasing the local merger rate would shift the allowed range for the beaming angle toward lower values so that the final coincident rate would be unchanged.
However, these estimates of the beaming angle should be considered with precaution due to the uncertainties associated with the intrinsic luminosity distribution derived by population synthesis from the small sample of observed sGRBs.

For ET the GW coincident rates are presented in Table~\ref{tab:ratesET} for a detector with a FOV of $4\pi$ sr, a duty cycle of 80\% and and infinite flux sensitivity. This is of course unrealistic but it gives an upper limit of the number of coincidences in $10-20$ years. If the sensitivity of GRB satellites do not improve in the next decade compared to {\it Swift}, the best one can expect is the {\it Swift} rate of 8 detections a year. This is much better that the rates predicted for ALV, but orders of magnitude below the ET potential. As a matter of consequence, we note that the construction of ET should be accompanied by the launch of a new generation of space observatories focussed on transient events.


\section{Constraints on the beaming angle}\label{sec:beaming_angle}

Coincident GW/GRB detections can help measuring, or at least putting strong constraints on the beaming angle of sGRBs \citep{Chen2013,Diez2011}. 
We propose here a very simple method to measure the average beaming factor and thus $<\cos \theta_B>$. 
The number of ALV coincident triggers will be probably too small to do a parameter estimation, thus we consider only the case of ET. We note that it is likely that a further enhancement of ALV, or a new detector before ET provide an intermediate case. 
For simplicity, we neglect NS-BH sources. 

In the redshift interval $z=0-0.2$, the GW efficiency of ET is almost 1 ($99.5 \%$ of the sources can be detected) and we assume the sensitivity of the satellite is good enough so that only a negligible fraction of all the sGRBs is below the flux threshold\footnote{with $F_{lim}=0.1-0.2$ ph~s$^{-1}$cm$^{-2}$, we have $99\%$ of the sources above the threshold  at a redshift of $z =0.2$ for instance}. The efficiency of coincident detections is then equal to the average beaming factor (times some factor due to the duty cycle and the FOV) and we can construct the estimator of $\Theta_B$:
\begin{equation}
\hat{\Theta}_B = (\varepsilon_{FOV} \times \varepsilon_{DC})^{-1} \frac{N^0_{\mathrm{cd}}}{N^0_{\mathrm{GW}}} ,
\end{equation}
\noindent where $N^0_{\mathrm{cd}}$ is the number of coincident detections and $N^0_{\mathrm{gw}}$ the number of GW detections alone, for $z$ in the bin $[0-0.2]$.  

\begin{figure}
\centering
\includegraphics[width=\columnwidth]{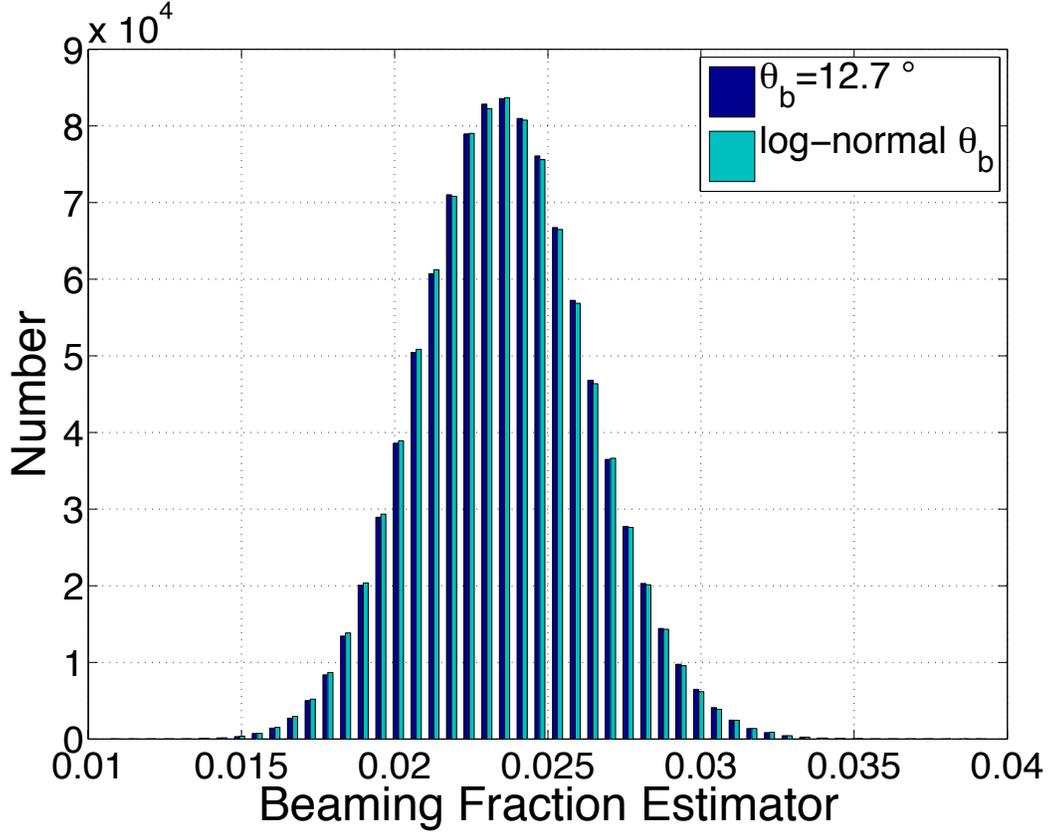}\\
\caption{\label{fig:histr} Histogram of $\hat{\Theta}_B$ from a sample of $10^5$ simulations assuming one year of observation, for the log-normal distribution with average value $\mu_{\log \theta_B}=2.0794$ and standard deviation $\sigma_{\log \theta_B}=0.69$ ($\theta_B$ in degrees) \citep{Goldstein11} compared to a fixed beaming angle $\theta_B=12.7^\circ$ giving the same average value of the beaming fraction $\Theta_B$. Here we have assumed a GRB satellite with a FOV of $4\pi$ sr, duty cycle of 100\% and infinite flux sensitivity..
}
\end{figure}

In order to test our method, we consider a population of BNS with local rate of $\dot{\rho}_c^o=0.1$ Mpc$^{-3}$ Myr$^{-1}$. Using the SFR of \citet{Hopkins06}, we obtain a total of $N^0_{\mathrm{GW}} \sim 3000$ GW over the operation lifetime of ET we assume to be 10 years.
We build the histogram of $\hat{\Theta}_B$ from a sample of $10^5$ simulations (Fig.~\ref{fig:histr}) and find that $\hat{\Theta}_B$ is a non-biased estimator of $\Theta_B$ (the average value $<\hat{\Theta}_B>$ converges to the true value) with an error that depends on the average number of coincident sources in one simulation, and thus on the SFR, the local rate, the beaming angle, the duty cycle and the FOV, and of course the time of observation. We confirm the result found in Section 3.4.2 that a distribution of the beaming angle does not affect the number of coincident detections, and thus the beaming factor estimator. However, when the distribution of $\theta_B$ is not known, we can only measure the average value of the cosine.

Table \ref{tab:params} gives the average value of $\hat{\Theta}_B$ and the standard deviation for the SFR of \citet{Hopkins06} ($<N^0_{\rm gw}> \sim 3000$) for $\varepsilon_{FOV}=\varepsilon_{DC}=1$ and fixed angles of 5$^{\circ}$, 10$^{\circ}$ and 15$^{\circ}$. 

For different combinations of local rates, SFR, duty cycle and FOV or time of observation $T$, one should simply rescale the standard deviation as:
\begin{equation}
\sigma' = \sqrt{\frac{3000}{<N_s>'}} \sigma , \\
\end{equation}
with: 
\begin{equation}
<N_s>'=T \times \varepsilon_{DC} \times \varepsilon_{FOV} \times \dot{\rho}_0 \times <N^0_{\rm gw}>,
\end{equation}
where $T$ is in year and $\dot{\rho}_0$ in Mpc$^{-3}$ Myr$^{-1}$.
For example the standard deviation should be multiplied by 3 for $\varepsilon_{FOV}$=0.15 and $\varepsilon_{DC}$=0.8. We have also studied the different distributions of $\theta_B$ used in III.D and confirmed this result.   

\section{Conclusion}\label{sec:conclusion} 

In this paper, we have presented Monte Carlo simulations of coincident detections  between gravitational wave and electromagnetic detectors. We have assumed that sGRBs could be powered by BNS or NS-BH coalescences and we have modeled the different selection effects of both GW and EM detectors. We have calculated a coincident efficiency taking into account the fact that the source inclination affect both the GW and GRB efficiency. 
Besides the beaming angle and the GRB satellite field of view, which are the most important effects, we have shown that the coincident sensitivity is limited by the GW detector horizon for the network of advanced detectors ALV, while for ET it is the GRB flux threshold that is the most important. For ALV the best GRB satellite will be the one with the largest field of view, independently of the flux sensitivity, but for ET, the most important gain will come from the improvement of the sensitivity.
In order to roughly estimate the sensitivity required for such an "optimized" sGRB detector used for the follow-up of ET, we have estimated that reducing the average flux limit to $F_{\mathrm{lim}} \sim 0.1-0.2$ ph~s$^{-1}$cm$^{-2}$ over the next decade, would allow 80\% of the sources at a redshift of $z \sim 1$ (corresponding to the end of the {\it plateau} of the coincident detection efficiency for BNS) to be above the threshold. The flux limit would have to be reduced to $F_{\mathrm{lim}} \sim 0.02-0.05$ ph~s$^{-1}$cm$^{-2}$ to obtain 80\% at $z=2$, where we have the peak of the distribution of the redshift, and where the difference between dark energy equation of states cosmological models is more visible.



Using a set of star formation rate models, we have calculated the coincident rate for different values of the beaming angle for ALV and ET. Our results predict a small number of coincident detections with ALV (less than one event per year for the {\it Swift} field of view of 1.4 sr and a duty cycle of 80 \%), in agreement with recent studies (see for instance \citet{Coward12,Petrillo13,Siellez14}). The observation of a sGRB counterpart, by increasing detection confidence, may allow for the reduction of the SNR threshold. We have shown that using a threshold of 8 rather than 12 increases the number of coincident detections by a factor of 3 for ALV. We have found a potential number of coincidences of $\sim 100-10000$  per year for ET assuming GRB satellites can reach the desired flux threshold and a maximum FOV of $\bf{4 \pi}$, but this number will reduce to a few events per year if the FOV and the sensitivity do not improve compared to {\it{Swift}}. 

Finally we have proposed an original method to estimate the mean jet opening angle of sGRBs. This method can be applied to ALV though with low sensitivity. The accuracy will improve as the sensitivity of the GW detectors enhance.

The coincident rates could slightly increase by considering the population of dynamical binaries that could have formed by captures in dense environment and that could be numerous at low redshift due to the long delay between formation and coalescence \citep{Wanderman14}. However we do not expect a big change and our findings emphasize the need of a dedicated, wide field of view, multi-wavelength follow-up of GW detections with a sensitivity increase by a factor 5 to 10 compared to current detectors.




\section*{Acknowledgments}

We acknowledge the use of public data from the \textit{Swift} data archive and Eric Howell and David Coward for helpful discussions.
DM acknowledges the PhD financial support from the Observatoire de la C$\hat{o}$te d'Azur and the PACA region. 




\newpage
\begin{deluxetable}{ccccc}
\tabletypesize{\scriptsize}
\tablecaption{Summary of the properties of the different electromagnetic detectors: the duty cycle DC, the field of view FOV in steradians, and the energy band in keV.\label{tab:detect}}
\tablewidth{0pt}
\tablehead{
\colhead{Mission} & \colhead{DC (\%)} & \colhead {FoV (sr)} &\colhead {Energy band (keV)}
}
\startdata
{\it Swift} & 80 & 1.4 & 15--150 \\
{\it \bf{Fermi}}-GBM & 80 & 9.5 & 8--30000 \\
{\it SVOM}    & 80 & 2      & 4--5000     \\
{\it LOFT}    & 80 & $\pi$  & 2--80     \\
\enddata
\end{deluxetable}

 \begin{deluxetable}{cccccccc}
\tabletypesize{\scriptsize}
\tablecaption{ BNS and NS-BH coincident detection rates (in units of yr$^{-1}$) for different values of the beaming angle, for ALV and the {\it Swift} GRB satellite with a FOV of 1.4 sr, a duty cycle of $80\%$, and a flux limit of $F_{lim}=1.5$ ph s$^{-1}$cm$^{-2}$. The range of values reflects the uncertainty on the SFR. The first line correspond to a signal-to-noise ratio thresholds of 12 and the second line to 8. We used the local rates of $\dot{\rho}_c^o=0.06$ Mpc$^{-3}$ Myr$^{-1}$ for BNS and 0.003 Mpc$^{-3}$ Myr$^{-1}$ for NS-BH, obtained recently by \citep{Dominik14} for the StarTrack standard model. For other local rates, one should multiply the values of Table~\ref{tab:rates} by $\dot{\rho}_c^o/0.06$ for BNS or $\dot{\rho}_c^o/0.003$ for NS-BH. The values in the last column indicates the GW detection rate.
The rates for the other satellites can be obtained by multiplying these results by FOV$/1.4$ sr, with the FOV given in Table~\ref{tab:detect}. 
\label{tab:rates}}
\tablewidth{0pt}
\tablehead{
 & \colhead{$5^\circ$ }& \colhead{$10^\circ$}& \colhead{$15^\circ$}&\colhead{ $20^\circ$} & \colhead{$30^\circ$} & \colhead{GW}  
}
\startdata
BNS  \\
$\rho_T=12$ & $0.004-0.005$ & $0.01-0.02$ & $0.03-0.04$ & $0.06-0.07$& $0.11-0.13$ & $2.5-3.0$\\
$\rho_T=8$ & $0.01-0.02$ & $0.05-0.06$ & $0.10-0.13$ & $0.18-0.23$ & $0.35-0.46$&\\
\hline
NS-BH  \\
$\rho_T=12$ &  $0.001-0.002$ & $0.006-0.008$ & $0.01-0.02$ & $0.02-0.03$ & $0.04-0.06$ & $1.5-2.0$\\
$\rho_T=8$ & $0.004-0.005$ & $0.01-0.02$ & $0.03-0.04$ & $0.05-0.08$ & $0.11-0.16$ & \\
\enddata
\end{deluxetable}

 \begin{deluxetable}{cccccccc}
\tabletypesize{\scriptsize}
\tablecaption{Same as Table ~\ref{tab:rates} for ET, assuming a GRB satellite with a FOV of $4\pi$ sr and infinite flux sensitivity, but accounting for the duty cycle of $80\%$ \label{tab:ratesET}}
\tablewidth{0pt}
\tablehead{
 & \colhead{$5^\circ$ }& \colhead{$10^\circ$}& \colhead{$15^\circ$}&\colhead{ $20^\circ$} & \colhead{$30^\circ$} & \colhead{GW}  
}
\startdata
BNS  &$(0.8-1.8) \times 10^2$ & $(3-7) \times 10^2$  & $(0.7-1.6) \times 10^3$  & $(1.3-2.8) \times 10^3$ & $(2.5-5.8)\times 10^3$ & $(0.6-1.5) \times 10^4$\\
\hline
NS-BH & $7-15$ & $27-61$ & $59-136$ & $104-239$ & $228-517$ & $(1.3-2.4) \times 10^3$\\
\enddata
\end{deluxetable}

\begin{deluxetable}{cccccc}
\tabletypesize{\scriptsize}
\tablecaption{Mean and standard deviation of the beaming angle estimator $\hat{\Theta}_B$ for the SFR of Hopkins ($<N^0_{\rm gw}> \sim 3000$), and assuming a GRB satellite with a FOV of $4\pi$ sr, duty cycle of 100\% and infinite flux sensitivity.)
\label{tab:params}}
\tablewidth{0pt}
\tablehead{
\colhead{$\theta_B$ ($^\circ$)}  & \colhead{$<\hat{\Theta}_B>$ }& \colhead{$\sigma_{\hat{\Theta}_B}$}& \colhead{error (\%)} \\
}
\startdata
5 & 0.0039 & 0.0011 & 29.3 \\
10 &  0.015 &  0.0022 & 14.5 \\
20 &  0.060 & 0.00042 & 7.23 \\
\enddata
\end{deluxetable}

\newpage

\newpage

\end{document}